\documentclass[twocolumn,floatfix,prb,aps,showpacs,superscriptaddress]{revtex4-1} 
\bibpunct{}{}{,}{n}{}{}

\usepackage{graphicx,epsf,amsmath,amssymb}
\usepackage{dcolumn}
\usepackage{bm,bbm}
\usepackage{enumerate}
\usepackage[usenames,dvipsnames,svgnames,table]{xcolor}

\usepackage{hyperref}
\hypersetup{
    bookmarks=false,         
    unicode=false,          
    pdftoolbar=true,        
    pdfmenubar=true,        
    pdffitwindow=false,     
    pdfstartview={FitH},    
    pdftitle={Superconductivity from Coulomb repulsion in 3D quadratic band touching Luttinger semimetals},    
    pdfauthor={UMontreal},     
    pdfsubject={Report},   
    pdfcreator={UMontreal},   
    pdfproducer={UMontreal}, 
    pdfkeywords= {superconductor} {repulsive} {transition temperature} 
    pdfnewwindow=true,      
    colorlinks=true,       
    linkcolor=blue,          
    citecolor=red,        
    filecolor=magenta,      
    urlcolor=cyan           
}

\begin{document}

\preprint{APS/123-QED}

\title{Superconductivity from Coulomb repulsion in three-dimensional quadratic band touching Luttinger semimetals}

\author{S. Tchoumakov}
\affiliation{D\'epartement de Physique, Universit\'e de Montr\'eal, Montr\'eal, Qu\'ebec, H3C 3J7, Canada}

\author{L. J. Godbout}
\affiliation{D\'epartement de Physique, Universit\'e de Montr\'eal, Montr\'eal, Qu\'ebec, H3C 3J7, Canada}

\author{W. Witczak-Krempa}
\affiliation{D\'epartement de Physique, Universit\'e de Montr\'eal, Montr\'eal, Qu\'ebec, H3C 3J7, Canada}
\affiliation{Centre de Recherches Math\'ematiques, Universit\'e de Montr\'eal; P.O. Box 6128, Centre-ville Station; Montr\'eal (Qu\'ebec), H3C 3J7, Canada}
\affiliation{Regroupement Qu\'eb\'ecois sur les Mat\'eriaux de Pointe (RQMP)}

\date{\today}

\begin{abstract}
We study superconductivity driven by screened Coulomb repulsion in three-dimensional Luttinger semimetals with a quadratic band touching and strong spin-orbit coupling. In these semimetals, the Cooper pairs are formed by spin-$3/2$ fermions with non-trivial wavefunctions. We numerically solve the linear Eliashberg equation to obtain the critical temperature of a singlet $s-$wave gap function as a function of doping, with account of spin-orbit and self-energy corrections. In order to understand the underlying mechanism of superconductivity, we compute the sensitivity of the critical temperature to changes in the dielectric function $\epsilon(i\Omega_n,q)$. We relate our results to the plasmon and Kohn-Luttinger mechanisms. Finally, we discuss the validity of our approach and compare our results to the litterature. We find good agreement with some bismuth-based half-Heuslers, such as YPtBi, and speculate on superconductivity in the iridate Pr$_2$Ir$_2$O$_7$. 
\end{abstract}


\maketitle

\section{Introduction}
The superconductivity of semiconducting materials has been experimentally studied since the 60s [\cite{reviewscsc}]. For most of these materials, such as diamond and silicon, experimental data and ab-initio calculations [\cite{diamond1,diamond2,silicon}] point towards a conventional pairing mechanism mediated by phonons [\cite{bcs,mahan}]. However, this picture does not seem to hold for some dilute semiconductors such as SrTiO$_3$~[\cite{behnia,gorkovA,gorkovB}], PbTe [\cite{buchauer}] and bismuth-based half-Heusler materials like YPtBi~[\cite{meinert}] where other pairing mechanisms have been suggested~[\cite{takada0,takada,ruhmanbismuth,savary}]. In various works it is proposed that YPtBi is a three-dimensional (3D) quadratic band-touching Luttinger semimetal~[\cite{halfheuslerbs1,luttinger,savary}], where the quasiparticles are characterized by a pseudospin $j = 3/2$ due to the strong spin-orbit coupling [\cite{savary,bitan,halfheuslerbs2,helicity}]. It is predicted that these semimetals exhibit topological surface states~[\cite{topolsm}]. Also, it was argued that Luttinger semimetals show non-Fermi liquid behaviour at small doping~[\cite{nfl1,nfl2,nfl3}], which makes them prime candidates to study novel phases arising from the interplay of spin-orbit coupling and electron interactions~[\cite{reviewwilliam}]. It was recently suggested that Luttinger semimetals, like YPtBi, constitute a platform for Cooper pairs of spin$-3/2$ fermions, including topological superconductivity~[\cite{halfheuslerbs2,savary,bitan,meinert,savary,topods,boettcher1,splinenode,boettcher2}].

In the present work we study singlet $s-$wave superconductivity in doped 3D Luttinger semimetals arising from the dynamically screened Coulomb repulsion between electrons. An analogous pairing mechanism was first proposed by Kohn and Luttinger in metals [\cite{kohnluttinger,chubukov}] who found that the screened Coulomb potential has attractive contributions that condense Cooper pairs with non-zero angular momentum ($\ell \neq 0$). This mechanism is of similar origin to the Friedel oscillations at $2k_F$, and usually leads to a small critical temperature. However, if one also considers the frequency dependence of the dielectric permittivity $\epsilon(i\Omega_n,q)$, within the random phase approximation (RPA), one finds a larger critical temperature~[\cite{takada0,takada}] that may account for the observed $T_c$ in SrTiO$_3$~[\cite{ruhmansto}]. In that case, contrary to the Kohn-Luttinger mechanism where the interaction is static, the critical temperature is obtained for a singlet $s-$wave order parameter and the gap function changes sign with frequency~[\cite{takada,ashcroft,ruhmanbismuth}]. The origin of the superconducting instability is then attributed to the electron-plasmon coupling because the screened Coulomb potential is negative for frequencies below the plasma frequency and above the region of electron-hole excitations~[\cite{takadavertex1}]. In this approach, most of the studies are based on a spin degenerate quadratic band model without spin-orbit coupling, which is well-suited for SrTiO$_3$~[\cite{takada,ruhmansto}] but not for Luttinger semimetals like YPtBi~[\cite{meinert}]. Indeed, it was recently shown that, compared to a single quadratic band, the interband coupling increases the long-range screening of the electric field and reduces the effective plasma frequency~[\cite{broerman,boettcher3,ourwork,amauri}]. This would weaken superconductivity within the aforementioned mechanism. This is without taking into account spin-orbit effects and the smaller self-energy corrections [\cite{ourwork}] that usually compete against superconductivity. 

In Sec.~\ref{sec:pairing} we discuss and solve the Eliashberg equation of a Luttinger semimetal with electron-electron repulsion and compare our results to the case of a single quadratic band. In Sec.~\ref{sec:sensitivity}, we discuss the influence of the frequency and wavevector dependence of the dielectric permittivity, $\epsilon(i\Omega_n,q)$, on the critical temperature. For this we compute the functional derivative $\delta T_c/\delta\epsilon(i\Omega_n,q)$ within a procedure similar to that used by Bergmann and Rainer in the 70s to discuss the effect of phonon softening in amorphous superconductors [\cite{bergmannrainer,allendynes}]. Finally, in Sec.~\ref{sec:discussion}, we compare our results to the experimental observations on superconducting Luttinger semimetals such as YPtBi, and we discuss the applicability of our theoretical description.

\section{Superconducting pairing}
\label{sec:pairing}
\begin{figure*}
    \centering
    \includegraphics[width = \textwidth]{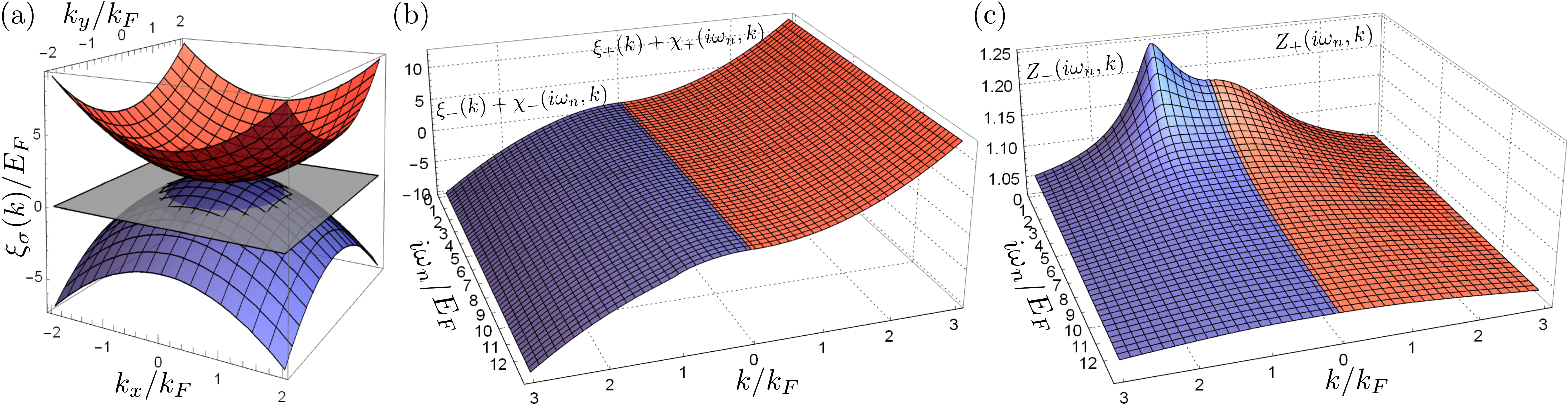}
    \caption{(a) Band-structure of the quadratic band touching Luttinger model; the gray plane is at the chemical potential. The upper (red) and lower (blue) bands are doubly degenerate, and we consider a singlet pairing within each band. (b,c) Behaviour of (b) $\xi_{\sigma}(i\omega_n,k) + \chi_{\sigma}(i\omega_n,k)$ and (c) $Z_{\sigma}(i\omega_n,k)$ for $r_s = 4$ with $\sigma = +$ in red and $-$ in blue. The Coulomb repulsion mostly affects the single-particle Hamiltonian at the Fermi energy (here, $\sigma = -$ and $(i\omega_n, k) = (0, k_F)$).}
    \label{fig:spenrg}
\end{figure*}
\subsection{Model}
The behaviour of non-interacting electrons at a quadratic band touching is described by the Luttinger Hamiltonian [\cite{luttinger}]
\begin{align}\label{eq:h0}
    \hat{H}_0 = \frac{\hbar^2}{2m} \left[ (\alpha_1  - 5\alpha_2/4) {\bf k}^2  + \alpha_2 \left({\bf k}\cdot \hat{{\bf J}}\right)^2 \right]  - \mu.
\end{align} 
We denote the band mass $m$ and the $j = 3/2$ total angular momentum operators $\hat{{\bf J}} = (\hat{J}_x,\hat{J}_y,\hat{J}_z)$. This model has inversion, rotation and time-reversal symmetry $\hat{\mathcal{T}}$~[\cite{murakami,savary,bitan}]. 
The Luttinger Hamiltonian describes four bands that meet quadratically at ${\bf k} = 0$, see Fig.~\ref{fig:spenrg}(a). The upper and lower bands are degenerate with energy $\varepsilon_{\sigma} = \sigma \hbar^2 k^2/(2 m_{\sigma})$ where $\sigma = \pm$. The upper and lower band masses, $m_{\pm} = m/(\alpha_2 \pm \alpha_1)$, are not necessarily the same. The eigenstates can be further decomposed in terms of Kramer partners $\lambda = \pm$, $|\sigma, \lambda, {\bf k}\rangle$ with $\lambda = \pm$, such that $|\sigma, +, {\bf k}\rangle = \hat{\mathcal{T}}|\sigma, -, {\bf k}\rangle$ [\cite{timereversal}]. It is convenient to introduce the projection operator $\hat{P}_{\sigma}({\bf k})$ on subband $\sigma$ with $\hat{P}_{\sigma}({\bf k}) = \frac12\left[ \hat{\mathbbm{1}} + \hat{H}_0({\bf k})/\xi_{\sigma}({\bf k})\right] = \sum_{\lambda = \pm} | \sigma, \lambda, {\bf k} \rangle\langle \sigma, \lambda, {\bf k} |$, where $\xi_{\sigma} = \varepsilon_{\sigma} - \mu$. This expression allows us to describe the eigenspinor overlap~[\cite{ffctr0,savary,broerman,boettcher3,ourwork,amauri}]
\begin{align}
    {\rm Tr}\!\!\left[\hat{P}_{\sigma_1}({\bf k}+{\bf q}) \hat{P}_{\sigma_2}({\bf k})\right] \!= \!\frac12\!\left\{ 2 + \sigma_1\sigma_2 \left[ 3 \cos^2(\theta_{{\bf k} + {\bf q}, {\bf k}}) - 1\right]\!\right\},
    \label{eq:projector}
\end{align}
which is central in the description of interband coupling. In the following we consider a particle-hole symmetric spectrum, where $\alpha_1 = 0$ and $\alpha_2 = 1$, with hole doping such that $E_F < 0$. We discuss the effect of particle-hole asymmetry in Sec.~\ref{sec:sensitivity}. 

The bare interaction between electrons is described by the Coulomb potential $V_0({ q}) = 4\pi e^2/(\epsilon^* q^2)$ where $\epsilon^*$ is the background dielectric permittivity. In second quantization, the full Hamiltonian is
\begin{align}
    \label{eq:ham}
    \hat{H} =& \sum_{\bf k}\hat{\psi}_{\bf k}^{\dagger} \hat{H}_{0}({\bf k}) \hat{\psi}_{\bf k}\\
    &+\frac{1}{2\mathcal{V}} \sum_{{s_1 s_2{\bf k}_1 {\bf k}_2,{\bf q}\neq 0}} V_0({ q}) \hat{\psi}^{\dagger}_{{\bf k}_1 + {\bf q}s_1} \hat{\psi}^{\dagger}_{{\bf k}_2 - {\bf q}s_2} \hat{\psi}_{{\bf k}_2s_2}\hat{\psi}_{{\bf k}_1s_1},\nonumber
\end{align}
where we introduce the fermionic annihilation operators $\hat{\psi}_{{\bf k}s} = \{\hat{\psi}_{{\bf k},3/2},\hat{\psi}_{{\bf k},1/2},\hat{\psi}_{{\bf k},-1/2},\hat{\psi}_{{\bf k},-3/2}\}$ of the $j=3/2$ representation.

In the following we set $\hbar = k_B = 1$. We write energies in units of the Fermi energy, $E_F$, and wavevectors in units of the Fermi wavevector, $k_F$. This choice of units allows us to write all expressions as a function of the Wigner-Seitz radius, $r_s = m e^2/(\alpha\epsilon^* k_F)$ with the constant $\alpha = (4/9\pi)^{1/3} \approx 0.52$, $k_F = (3\pi^2 n)^{1/3}$ and where $E_F = -k_F^2/2m < 0$ is in the bottom band. The band structure is particle-hole symmetric and because we consider $s-$wave pairing, our observations are independent on the sign of the Fermi energy [\cite{savary}].

\subsection{Eliashberg equation}

The Eliashberg equation [\cite{eliashberg}] of the $j = 3/2$ electrons has recently been discussed in Refs.~[\cite{savary,bitan}]. They describe the various pairing channels of Luttinger semimetals and the corresponding coupling strength due to polar optical phonons~[\cite{savary}]. However, the electronic polarization is only accounted for approximatively and in the present section we consider how it can be responsible for pseudo-spin-singlet superconductivity. Here, the pseudo-spin refers to the two Kramer partners within a band. We consider that pairing occurs through the screened Coulomb potential
\begin{align}\label{eq:epol}
    V(i\Omega, {\bf q}) = \left.V_{0}({q})\right/{\epsilon_{\rm RPA}(i\Omega, {\bf q})},
\end{align}
where $\epsilon_{\rm RPA}(i\Omega, {\bf q})$ is the dielectric permittivity in the random phase approximation. This expression has been computed at zero temperature in Refs.~[\cite{broerman,boettcher3,ourwork,amauri}] but it should not strongly differ from that at the critical temperature, $T_c$, since $T_c \ll E_F$, as we shall see. In a Luttinger semimetal, screening at small wavevectors is stronger than for a single quadratic band [\cite{ourwork}] which leads to a smaller plasma frequency and a reduced renormalization of the quasiparticle properties. 

We rewrite the Hamiltonian \eqref{eq:ham} in terms of fermionic operators associated with the eigenstates of $\hat{H}_0$, $\hat{c}_{{\bf k}\sigma\lambda} = \sum_{s} \langle {\bf k}, \sigma, \lambda| \psi_{{\bf k}s} \rangle \hat{\psi}_{{\bf k}s}$, where $| \psi_{{\bf k}s} \rangle$ are the states in the original basis. We consider the normal and anomalous Green's functions
\begin{align}
    &G_{\sigma\lambda}(i\omega_n,{\bf k}) = - \int_{0}^{1/T} d\tau \langle \hat{T}_{\tau} \hat{c}_{{\bf k}\sigma\lambda}(\tau) \hat{c}_{{\bf k}\sigma\lambda}^{\dagger}(0) \rangle e^{i\omega_n \tau},\\
    &F_{\sigma}(i\omega_n,{\bf k}) = - \int_{0}^{1/T} d\tau \langle \hat{T}_{\tau} \hat{c}_{{\bf k}\sigma +}(\tau) \hat{c}_{-{\bf k}\sigma -}(0) \rangle e^{i\omega_n \tau}.
\end{align}
Here we denote the fermionic Matsubara frequencies $\omega_n = (2n+1)\pi T$ with $T$ the temperature and $n$ an integer, the ordering operator in imaginary time $\hat{T}_{\tau}$, the thermal average $\langle \cdots \rangle = {\rm Tr}( e^{-\beta\hat{H}} \cdots)/{\rm Tr}( e^{-\beta\hat{H}})$ with $\hat{c}_{{\bf k}\sigma\lambda}(\tau) = e^{\tau \hat{H}}\hat{c}_{{\bf k}\sigma\lambda}e^{-\tau \hat{H}}$. In Appendix~\ref{app:elshberg} we derive the equations of motion of the normal and anomalous Green's function. 

The self-energy $\Sigma_{\sigma}(i\omega_n, {\bf k})$ of a quadratic band touching Luttinger semimetal in the normal state was numerically obtained in Ref.~[\cite{ourwork}]. These expressions are independent of $\lambda$ and we can thus omit the index $\lambda$ for the normal Green's function (see Appendix~\ref{sec:app:selfe}). The Green's functions are $G_{\sigma}(i\omega_n,{\bf k}) = [G^{(0)}_{\sigma}(i\omega_n,{\bf k})^{-1} - \Sigma_{\sigma}(i\omega_n, {\bf k})]^{-1}$ where the bare Green's functions are $G^{(0)}_{\sigma}(i\omega_n,{\bf k}) = (-i\omega_n + \xi_{\sigma}({\bf k}))^{-1}$. The time-reversal symmetry allows the decomposition of the self-energy over two functions, $\chi_{\sigma}$ and $Z_{\sigma}$, that are real and even in $\omega_n$
\begin{align}\label{eq:self}
    \Sigma_{\sigma}(i\omega_n,{\bf k}) = \chi_{\sigma}(i\omega_n,{\bf k}) + i \omega_n(1 - Z_{\sigma}(i \omega_n,{\bf k})).
\end{align} 
Also, because of rotational symmetry, the dependence of the self-energy on wavevectors is only through $k = |{\bf k}|$. In Fig.~\ref{fig:spenrg}(b,c) we illustrate the typical behaviour of $\chi_{\pm}$ and $Z_{\pm}$. This behaviour of the self-energy is qualitatively the same for all values of $r_s$. 
The value of $Z_{\sigma}^{-1}(i\omega_n, k)$ is the quasiparticle weight and is peaked at the Fermi surface, at $(\sigma, i\omega_n, k) = (-, \pi T, k_F)$. It decreases to unity away from the Fermi surface, as seen in Fig.~\ref{fig:spenrg}(c).

The two anomalous Green's functions $F_{\pm}(i\omega_n,{\bf k})$ describe two singlet Cooper pairs, one for each subband. The critical temperature of a superconductor is often determined with the linearized Eliashberg equation in terms of the gap functions [\cite{kohnluttinger,takada,ashcroft,ruhmanbismuth}]
\begin{align}
    \label{eq:stdphi}
    \phi_{\pm}(i\omega_n, {\bf k}) \equiv G_{\pm}^{-1}(i\omega_n, {\bf k}) G_{\pm}^{-1}(-i\omega_n, {\bf k}) F_{\pm}(i\omega_n, {\bf k}),
\end{align}
that in the present situation can be decomposed over spherical harmonics $\phi_{\sigma}(i\omega_n, {\bf k}) = \sum_{\ell m} \phi_{\ell\sigma}(i\omega_n,k) Y_{\ell m}(\theta, \phi)$ as in Refs.~[\cite{kohnluttinger,takada}]. In the present work, we instead consider this self-consistent equation in terms of the barred gap function $\bar{\phi}_{\ell\pm}$
\begin{align}
    \label{eq:symphi}
    \bar{\phi}_{\ell\pm}(i\omega_n,k) \equiv k F_{\ell\pm}(i\omega_n,k),
\end{align} 
such that the linearized Eliashberg equation becomes a \emph{symmetric} eigenvalue equation of the form $\rho_{} \bar{\phi}_{\ell} = \bar{S}_{\ell}\bar{\phi}_{\ell}$~(see Appendix~\ref{app:elshberg}) :
\begin{align}
    \label{eq:eshsym}
    &\rho_{} \bar{\phi}_{\ell\sigma_1}(i\omega_{n_1},k_1) = - \sum_{\omega_{n_2}\sigma_2} \int dk_2~ \{ I_{\ell\sigma_1\sigma_2}(i\omega_{n_1}, k_1 ; i\omega_{n_2}, k_2)\nonumber\\
    & + \delta(k_1\!-\!k_2) \delta_{n_1 n_2} \delta_{\sigma_1\sigma_2} T^{-1} K_{\sigma_2}(i\omega_{n_2}, k_2) \} \bar{\phi}_{\ell\sigma_2}(i\omega_{n_2},\!k_2),
\end{align}
where the largest eigenvalue $\rho_{\rm max}(T)$ vanishes at the critical temperature $T_c$, \emph{i.e.} $\rho_{\rm max}(T_c) = 0$.

In this work we only study the $s-$wave ($\ell = 0$) pairing channel, even in Matsubara frequencies. The electron pairing interaction in Eq.~\eqref{eq:eshsym} then becomes
\begin{align}
    \label{eq:iav}
    I_{0\sigma_1\sigma_2} = &\int_{|k_1-k_2|}^{k_1+k_2} dq \frac{q V_0(q) N_0}{\epsilon(i(\omega_{n_1} - \omega_{n_2}),q)}\\
    &\times \frac14\left\{ 2 + \sigma_1\sigma_2\left[ 3 \left(\frac{k_1^2 +k_2^2 - q^2}{2k_1k_2}\right)^2 - 1 \right] \right\} \nonumber,
\end{align}
where $N_0 = 1/(4\pi^2)$ is the dimensionless density of states per band at the Fermi surface. The term in curly braces arises from the eigenspinor overlap in Eq.~\eqref{eq:projector}. In Eq.~\eqref{eq:eshsym}, the pairing potential competes with the kinetic contribution 
\begin{align}
    \label{eq:Ks}
    K_{\sigma} = (\omega_{n} Z_{\sigma}(i\omega_n,k))^2 + (\xi_{\sigma}(k) + \chi_{\sigma}(i\omega_n,k))^2.
\end{align}
This expression contains the renormalized single-particle Green's function, which we illustrate in Fig.~\ref{fig:spenrg}(b,c). This depairing contribution is diagonal in Eq.~\eqref{eq:eshsym}, it is also positive and has a minimum at the Fermi surface.

\begin{figure}[t]
    \centering
    \includegraphics[width = \columnwidth]{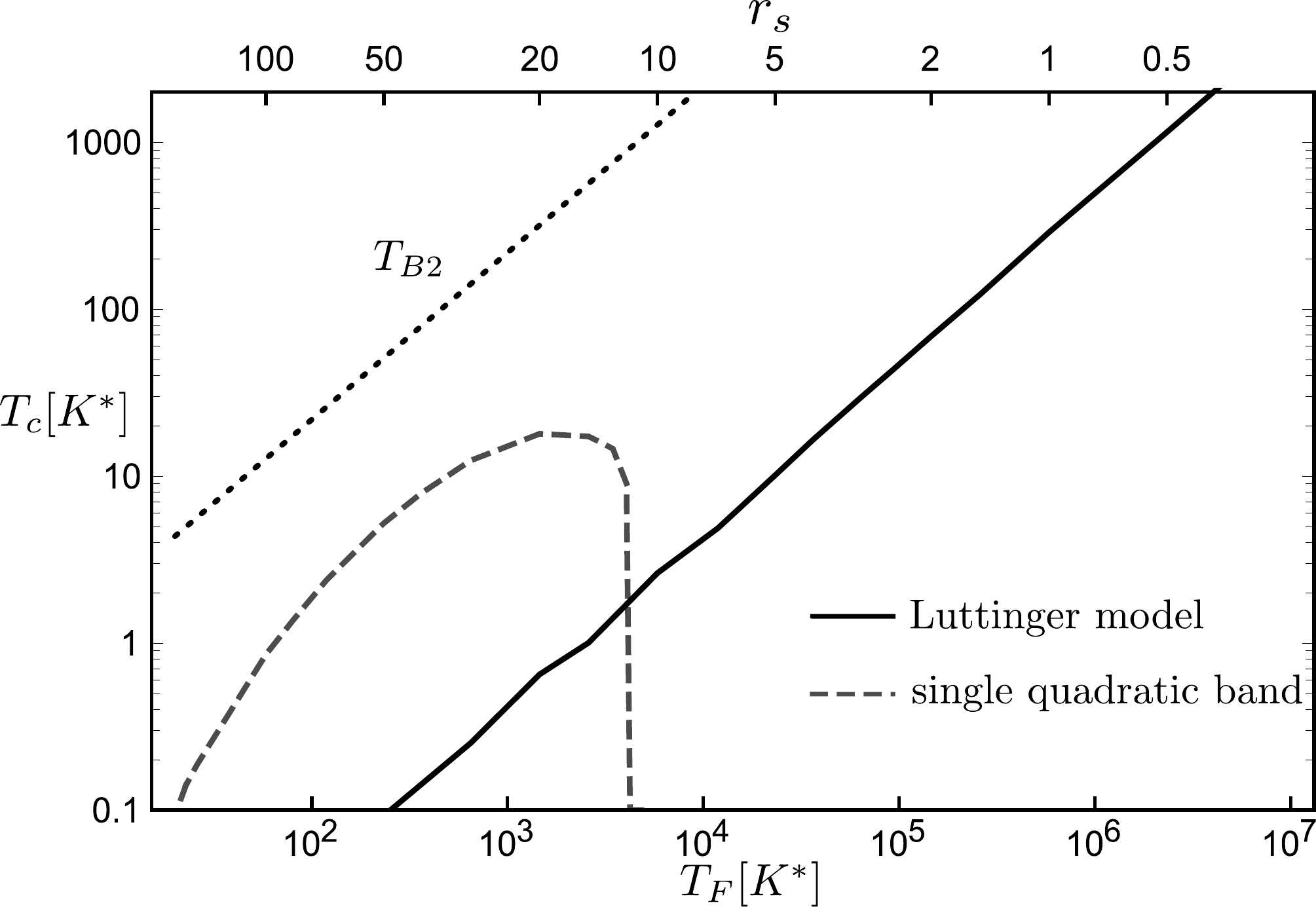}
    \caption{Critical temperature, $T_{c}$, in units of ${\rm K}^* = m/(m_e\epsilon^{*2}) {\rm K}$, as a function of the Fermi temperature and the Wigner-Seitz radius, for the Luttinger semimetal (plain) and for the single quadratic band~[\cite{takada}] (gray, dashed). As a reference, we show the Bose-Einstein condensation temperature $T_{B2}$ for free bosons with effective mass $2m$ and density $n/2$.}
    \label{fig:tc12}
\end{figure}

The linear Eliashberg equation in Eq.~\eqref{eq:eshsym} is symmetric for the canonical scalar product on $\sigma = \pm$, $i\omega_n$ and $k$, which is useful to reduce the number of numerical operations to solve it. It also allows the use of variational properties of symmetric equations. For example, for any test function $\phi_{}^{t}$, one has $\rho_{\rm max}(T) > \rho^{t}(T) = \phi_{}^{t}\cdot \hat{S}_{}\phi_{}^{t}/(\phi_{}^{t}\cdot \phi_{}^{t})$. The numerical determination of the critical temperature $T^{t}_c$ is thus bounded from above by its exact value, $T_c > T_{c}^{t}$. A similar bound is discussed in the context of phonon-mediated superconductivity in Ref.~[\cite{allendynes}]. Another use of Eq.~\eqref{eq:eshsym} is in the determination of the sensitivity of the critical temperature to changes in the dielectric permittivity, $\epsilon(i\Omega_n,q)$, that is the functional derivative $\delta T_c/\delta \epsilon(i\Omega_n,q)$, through the Hellmann-Feynman theorem. A similar calculation is performed in Ref.~[\cite{bergmannrainer}] to obtain the sensitivity of the critical temperature to the density of states of phonons. We discuss this last point in Sec.~\ref{sec:sensitivity}. 

\subsection{Numerical solution}
\label{subsec:num}

We solve the symmetric linear Eliashberg equation \eqref{eq:eshsym} numerically by decomposing $\bar{\phi}$ on its components ${\Delta}_{\pm,sd}$ on a grid $\{\nu_s\}_{s \in [1,N_1]}$ of imaginary frequencies and $\{k_d\}_{d\in[1,N_2]}$ of wavevectors
\begin{align}\label{eq:decompos}
    \bar{\phi}_{\pm}(i\omega_{n},k) = \sum_{\substack{s \in [1,N_1]\\ d\in[1,N_2]}} {\Delta}_{\pm,sd}~\Pi^{(1)}_{s}(\omega_n) \Pi^{(2)}_{d}(k).
\end{align}
In this decomposition we use the normalized rectangular functions $\Pi^{(1)}_{s < N_1}(\omega_n)$ and $\Pi^{(2)}_{d < N_2}(k)$ that are respectively constant in the intervals $\omega_n \in [\nu_s, \nu_{s+1}]$ and $k \in [k_d, k_{d+1}]$, and zero otherwise. The asymptotic behaviour of the linear Eliashberg equation enforces that for $\omega_n \gg 1$, $\bar{\phi}_{\ell}(i\omega_n,k) \sim 1/\omega_n^2$ and that for $k \gg 1$, $\bar{\phi}_{\ell}(i\omega_n,k) \sim 1/k^5$~[\cite{rspseudopot}]. We thus complete the grid for $\omega_n \geq \omega_{N_1}$ and for $k \geq k_{N_2}$ with the normalized asymptotic functions $\Pi^{(1)}_{s = N_1}(\omega_n) = \sqrt{6\pi T \omega_{N_1}^3}/\omega_n^2$ and $\Pi^{(2)}_{d = N_2}(k) = \sqrt{9 k_{N_2}^{9}}/k^5$. The grid in frequency and in wavevectors is refined to converge to a stable solution (see Appendix~\ref{app:asymptotic}).

\begin{figure}
    \centering
    \includegraphics[width = \columnwidth]{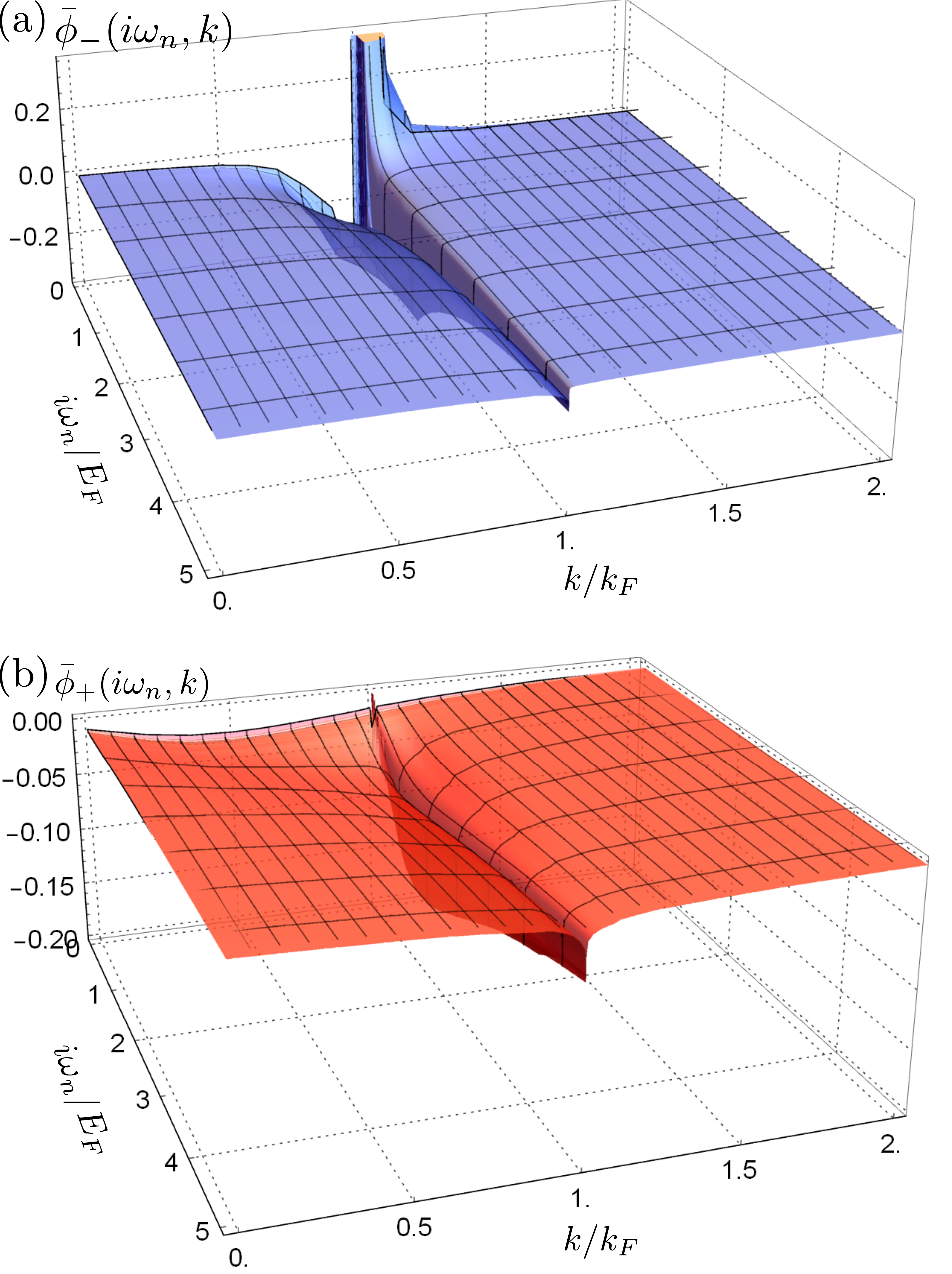}
    \caption{
    Gap functions $\bar{\phi}_{\pm}(i\omega_n,k)$ \eqref{eq:symphi} on the (a) lower and (b) upper band for $\ell = 0$ ($s-$wave) and $r_s = 5$, they are normalized such that $\bar{\phi}_{-}(i\pi T_c,k_F) = 1/(\pi T_c)^2$. The peak at $k = k_F$ indicates that pairing mostly occurs at the Fermi surface.}
    \label{fig:gap12}
\end{figure}

With this decomposition, Eq.~\eqref{eq:eshsym} becomes the matrix eigenvalue equation
\begin{align}
     \rho_{} {\Delta}_{\sigma_1 s_1d_1} = \sum_{\sigma_2 = \pm}\sum_{s_2=1}^{N_1} \sum_{d_2=1}^{N_2} {S}_{\sigma_1s_1d_1,\sigma_2 s_2d_2} {\Delta}_{\sigma_2 s_2 d_2},
     \label{eq:elshsymnum}
\end{align}
where the matrix components are
\begin{align}
    &{S}_{\sigma_1s_1d_1,\sigma_2 s_2d_2} = \frac{-1}{\sqrt{N_{s_1}(T)N_{s_2}(T)}} \!\sum_{\sigma_2}\!\!\sum_{\substack{\nu_{s_1} \leq \omega_{n_1} < \nu_{s_1+1} \\ \nu_{s_2} \leq \omega_{n_2} < \nu_{s_2+1}}}
    \label{eq:snume}\\
    &\int_{k_{d_1}}^{k_{d_1+1}} \!\!\!\!\frac{dk_1}{\Delta k_{d_1}} \!\!\int_{k_{d_{2}}}^{k_{d_{2}+1}}\!\!\!\! \frac{dk_2}{\Delta k_{d_2}}( I_{0\sigma_1\sigma_2}\!\!- \delta_{k_1k_2} \delta_{n_1 n_2} \delta_{\sigma_1\sigma_2}T^{-1} K_{\sigma_2} )\nonumber,
\end{align}
where the function $N_s(T)$ counts the number of Matsubara frequencies in the interval $[\nu_s, \nu_{s+1}]$. 
The discrete summations over Matsubara frequencies, for $\omega_{n_1},\omega_{n_2} \leq \omega_{N_1}$, are obtained with a linear interpolation of $I_{0}$ and $K$ over the grid of frequencies $\{\nu_s\}_{s \in [1,N_1]}$. We convert this summation to an integral for $\omega_n \geq \omega_{N_1}$, $\sum_{\omega_n} \approx \frac{1}{2\pi T}\int_{\omega_{N_1}}^{\infty} d\omega$. Note that a decomposition similar to~\eqref{eq:snume} is performed in Ref.~[\cite{takada}] but the normalization factors do not appear explicitly. There, a sum and an integral are approximated with a Riemann summation that absorbs the normalization factors without affecting the eigenvalues. 

We compute the critical temperature $T_c$ by solving the equation $\rho(T = T_c) = 0$ for different values of the Wigner-Seitz radius. We report our results in Fig.~\ref{fig:tc12} with temperatures given in units of ${\rm K}^* = (m/m_e)/\epsilon^{*2} {\rm K}$. We also show the results for a single quadratic band~[\cite{takada}], which we have reproduced with the aforementioned methodology. In contrast to a single quadratic band, the superconductivity of a Luttinger semimetal persists in the regime of small Wigner-Seitz radii. We find a solution down to $r_s = 0.01$ and the critical temperature drops below this limit. 
We observe that the critical temperature scales linearly, with $T_c/T_F \approx 4.4(4)\times 10^{-4}$. This can be compared to the ratio found for a single quadratic band at large $r_s$ where $T_c/T_F \approx 0.02$~[\cite{takada}]. 

\subsection{Structure of the gap function}
\label{subsec:sgap}

In Figs.~\ref{fig:gap12}(a,b) we show the components $\bar{\phi}_{\pm}$ of the gap function \eqref{eq:symphi}. The gap function has a larger weight close to the Fermi surface, for $\sigma = -$, $k = k_F$ and small $\omega_n$. At $k = k_F$ the gap function $\bar{\phi}_-$ changes sign as a consequence of the repulsive nature of the Coulomb potential and this change of sign is also present for $\phi_{-}$ in~\eqref{eq:stdphi}. We note that, at fixed $k$, the gap function $\bar{\phi}_-$ does not change sign as a function of the imaginary frequency in contrast with the gap function in a single quadratic band~[\cite{takada,ashcroft}].
Interestingly, the contribution of the gap function on the upper band, $\bar{\phi}_+$, is non-negligible away from $(\omega_n,k) = (0,k_F)$ and the sign of $\bar{\phi}_+$ is opposite to $\bar{\phi}_-$ for $k > k_F$. The opposite sign of the gap function on the two bands reminds observations of opposite $s-$wave order parameters on electron and hole bands in FeSe~[\cite{fese1,fese2}]. In Fig.~\ref{fig:gapratio} in Appendix~\ref{app:asymptotic} we plot the ratio of $\bar{\phi}_{+}$ and $\bar{\phi}_{-}$ to further show the importance of $\bar{\phi}_+$. In absence of the contribution on the upper band, \emph{i.e.} when setting $\bar{\phi}_+ = 0$, we do not find a critical temperature above the lowest temperature, $T/T_F \approx 10^{-5}$, achievable with our numerical accuracy (see Appendix~\ref{app:asymptotic}). This indicates the importance of interband coupling due to the spin-orbit interaction in the present mechanism of superconductivity. 

\subsection{Superconductivity from polar phonon screening}
\label{subsec:phon}

The superconductivity of Luttinger semimetals mediated by the Coulomb repulsion was first discussed in Ref.~[\cite{savary}] in the context of the superconductivity of YPtBi. There, corrections to the electronic self-energy are neglected and the Coulomb repulsion is mediated by the dynamic polarization of optical phonon modes :
\begin{align}
    \label{eq:epssav}
    \epsilon(i\Omega, q) = \epsilon_{\infty} + \frac{\epsilon_0 - \epsilon_{\infty}}{1 + \Omega^2/\omega_T^2} + \frac{q_{TF}^2}{q^2}.
\end{align}
This dielectric function accounts for the screening by polar phonon modes of frequency $\omega_T$ [\cite{lst}], and the electronic screening is described by the Thomas-Fermi wavevector $q_{TF}^2 = 2 \epsilon_{\infty} \alpha r_s/\pi$ [\cite{mahan}]. The authors in [\cite{savary}] compute the critical temperature with a pseudo-potential approach [\cite{pseudopot,rspseudopot}] where the characteristic frequency $\omega_c$ equals the phonon frequency $\omega_T$. In this kind of approximation, emphasis is put on the retardation of the pairing potential, that is on its frequency dependence below and above a characteristic frequency $\omega_c$. Using their approximate expression for $T_c$ and sensible values for YPtBi [\cite{valphon}] we evaluate $T_c < 0.01$~K for $s-$wave pairing in YPtBi, which corresponds to $T_c/T_F < 4\times 10^{-6}$. This is consistent with our full numerical solution of the Eliashberg equation with the dielectric function \eqref{eq:epssav}, with and without the self-energy corrections. Indeed, we do not find any solution above the lowest temperature achievable with our numerical accuracy, which corresponds to $T/T_F \approx 10^{-5}$ (see Appendix~\ref{app:asymptotic}). 

~~

The previous calculations show that the superconductivity from the Coulomb repulsion in Luttinger semimetals strongly relies on the interband coupling and on the screening mechanism. At large $r_s$, the critical temperature is much smaller than for a single quadratic band but extends to small values of the Wigner-Seitz radius, down to $r_s \approx 0.01$, due to the interband coupling. Even so, we compute a critical temperature about two orders of magnitude larger than with optical phonon modes [\cite{savary}]. In the following we make use of the symmetry in Eq.~\eqref{eq:eshsym} and explore how each component ($i\Omega_n$, ${\bf q}$) of $\epsilon_{\rm RPA}(i\Omega_n,{\bf q})$ in Eq.~\eqref{eq:epol} affects the critical temperature.

\section{Sensitivity of the critical temperature to screening}
\label{sec:sensitivity}

The observed differences between the critical temperature of a single quadratic band and of the quadratic band touching Luttinger semimetal come from the wavefunction overlap and the effect of interband coupling on the screening function $\epsilon_{\rm RPA}(i\Omega,{\bf q})$~\eqref{eq:epol}. They both reduce the pairing potential, leading to a smaller critical temperature compared to a single quadratic band in the regime of large $r_s$. However, the larger screening of the Coulomb potential also reduces the importance of the self-energy and allows for the observation of superconductivity at smaller values of the Wigner-Seitz radius. In order to further discuss the underlying mechanisms of superconductivity, one can explore the sensitivity of the critical temperature to changes in the dielectric permittivity.

\begin{figure*}[htb]
    \centering
    \includegraphics[width=\textwidth]{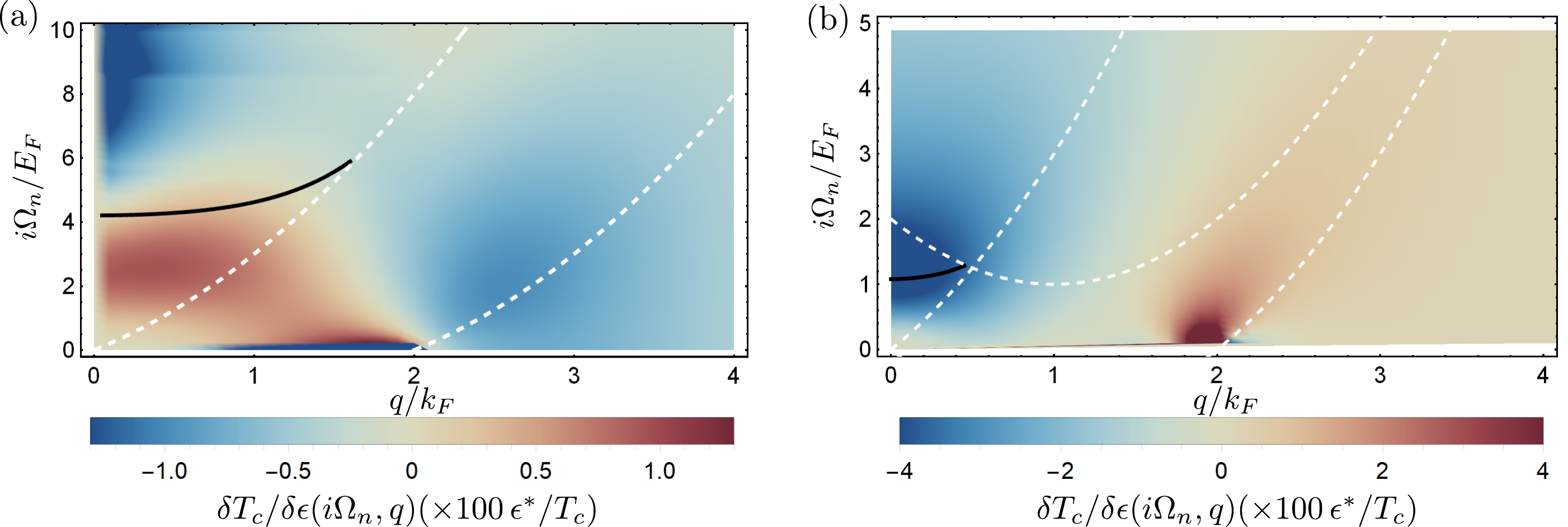}
    \caption{Functional derivative ${\delta T_c}/{\delta \epsilon(i\Omega_n,q)}$ of the critical temperature over the dielectric permittivity at $r_s = 20$ in percentage of $T_c/\epsilon^*$ (a) for a single quadratic band and (b) for the quadratic band touching Luttinger model. The black line is the dispersion relation of plasmons and the white dashed lines are the branches of the particle-hole excitation diagram. These lines are a guide to the eye, since they are computed for real frequencies. These figures indicate that the critical temperature is mostly sensitive to the long-range screening from plasma oscillations, and from the static scattering at $2k_F$.}
    \label{fig:sensitivity}
\end{figure*}

As illustrated in the previous section, the critical temperature of a superconductor relies on an integral equation \eqref{eq:eshsym} over all the components of the dielectric permittivity $\epsilon(i\Omega_n,q)$. If one changes $\epsilon(i\Omega_n,q)$ by $\delta \epsilon(i\Omega_n,q)$ then the change in the critical temperature $\Delta T_c$ is
\begin{align}\label{eq:defdtc}
    \Delta T_c = 2\pi T\sum_{\Omega_n} \int dq~ \frac{\delta T_c}{\delta \epsilon(i\Omega_n,q)} \delta \epsilon(i\Omega_n,q) .
\end{align}
The functional derivative ${\delta T_c}/{\delta \epsilon(i\Omega_n,q)}$ measures the sensitivity of the critical temperature to screening, and it is large for components $(i\Omega_n,q)$ responsible for the superconducting condensation. A similar quantity is defined in the context of the electron-phonon mechanism of superconductivity [\onlinecite{bergmannrainer}] to discuss the optimal phonon spectrum for the largest critical temperature [\cite{allendynes, leavens}]. Note that in Eq.~\eqref{eq:defdtc} we consider the sensitivity of the critical temperature to the dielectric permittivity for imaginary frequencies. As such, its physical interpretation is not straightforward but it is related to the behaviour at real frequencies by the continuation $i\Omega \rightarrow \omega + i 0^+$ and thus shows similar characteristic behaviour. We account for this aspect in our discussion.

Since $ \rho_{}(T_c) = 0$, the functional derivative ${\delta T_c}/{\delta \epsilon(i\Omega_n,q)}$ satisfies the relation
\begin{align}
    \label{eq:tride}
    \frac{\delta T_c}{\delta \epsilon(i\Omega_n,q)} = - \left.\frac{\delta  \rho_{}}{\delta \epsilon(i\Omega_n,q)}\right|_{T=T_c}\left/\frac{\partial  \rho_{}}{\partial T}\right|_{T=T_c},
\end{align}
which simplifies its numerical evaluation. Here $\rho$ is the maximal eigenvalue of the Eliashberg equation \eqref{eq:elshsymnum} from which $\partial\rho/\partial T|_{T=T_c}$ can be numerically approximated. Also, because Eq.~\eqref{eq:eshsym} is symmetric, we use the Hellmann-Feynman theorem to write
\begin{align}
    \frac{\delta  \rho_{}}{\delta \epsilon(i\Omega_n,q)} = \frac{\bar{\phi}_{}\cdot \left( {\delta {S}_{}}/{\delta \epsilon(i\Omega_n,q)}\right)\bar{\phi}_{}}{\bar{\phi}_{}\cdot\bar{\phi}_{}},
\end{align}
where $\bar{\phi}_{}$ is the eigenvector corresponding to $\rho_{\rm max}(T = T_c) = 0$. For a normalized eigenvector the expression is
\begin{widetext}
    \begin{align}
        &\frac{\delta \rho_{}}{\delta \epsilon(i\Omega_p,q)}\bigg|_{T_c} = \frac{V_0(q)N_0}{\epsilon^2(i\Omega_p,q)}\sum_{\substack{\omega_{n_1}\omega_{n_2} >0\\ \sigma_1\sigma_2}}\int_{0}^{\infty} dk_1 dk_2 \bar{\phi}_{\sigma_1}(i\omega_{n_1},k_1)\bar{\phi}_{\sigma_2}(i\omega_{n_2},k_2)
        \label{eq:drhodeps}\\
        &~~~~~~\times\bigg\{ \frac14\left( 2 + \sigma_1 \sigma_2 \left[ 3 \left(\frac{k_1^2 +k_2^2 - q^2}{2k_1k_2}\right)^2 - 1 \right] \right) q\Theta(k_1+k_2-q)\Theta(q-|k_1-k_2|) \delta_{p,n_1-n_2} \nonumber\\
        &- \delta(k_1 - k_2) \delta_{\sigma_1 n_1,\sigma_2 n_2} 2 q^2\left(\omega_{n_2} Z_{\sigma_2}(i\omega_{n_2},k_2) \bar{G}_{\sigma_2}^{(2)}(i(\omega_{n_2} -\Omega_p),k_2, q) - (\xi_{\sigma_2}(k_2) + \chi_{\sigma_2}(i\omega_{n_2},k_2)) \bar{G}_{\sigma_2}^{(1)}(i(\omega_{n_2} -\Omega_p),k_2, q)\right) \bigg\},\nonumber\\
        &+ ( i\Omega_p \rightarrow -i\Omega_p )\nonumber
    \end{align}
\end{widetext}
where $\bar{G}^{(1)}_{\pm}$ and $\bar{G}^{(2)}_{\pm}$ are coupling factors computed in Appendix~\ref{app:gtheta} and $\Theta(x)$ is the Heaviside step function. The last term in Eq.~\eqref{eq:drhodeps} appears because we consider a gap function even in frequency ($\ell = 0$) and because $\epsilon(-i\Omega_p,q) = \epsilon(i\Omega_p,q)$ [\cite{ferrel}]. This functional derivative includes the effect of the dielectric permittivity on both the pairing potential and the single-particle self-energy of a Luttinger semimetal. A similar expression is obtained for a single quadratic band by removing the wavefunction overlap. We compute the derivative \eqref{eq:drhodeps} numerically with a procedure similar to that presented in Sec.~\ref{subsec:num}. 

In Fig.~\ref{fig:sensitivity} we show the functional derivative ${\delta T_c}/{\delta \epsilon(i\Omega_n,q)}$ in percentage of $T_c/\epsilon^*$ for (a) the single quadratic band and for (b) the quadratic band touching Luttinger model. The screening mechanisms that act positively (negatively) on the critical temperature are in red (blue). For example, the change of sign of ${\delta T_c}/{\delta \epsilon(i\Omega_n,q)}$ from positive below $q = 2k_F$ to negative above it, indicates that increasing the discontinuity in the static dielectric function at $2k_F$ increases the critical temperature. This is a signature of the Kohn-Luttinger mechanism of superconductivity [\cite{kohnluttinger}] which relies on the discontinuity of screening at $2k_F$. This mechanisms happens for the two bandstructures under study but it becomes negligible for larger values of $r_s$ for a single quadratic band, where the plasmon mechanism of superconductivity dominates [\cite{takada}]. The signature of the plasmon mechanism appears in Fig.~\ref{fig:sensitivity}(a,b) for small values of $q$ and at finite Matsubara frequencies. For a single quadratic band, in Fig.~\ref{fig:sensitivity}(a), the sign of ${\delta T_c}/{\delta \epsilon(i\Omega_n,q)}$ is consistent with an increase in the critical temperature from an increase in the plasma frequency $\omega_p$ in the optical permittivity $\epsilon_{0}(i\Omega_n, q = 0) = 1 + \omega_p^2/\Omega_n^2$~[\cite{mahan}]. On the other side, in a Luttinger semimetal, in Fig.~\ref{fig:sensitivity}(b), the sign of ${\delta T_c}/{\delta \epsilon(i\Omega_n,q)}$ is opposite, with a strong negative contribution above the plasma frequency. We associate this behaviour to the interband coupling that strongly increases the dielectric permittivity at the onset of interband transitions, at $\omega = 2 E_F$, and is responsible for a decrease in the plasma frequency of Luttinger semimetals [\cite{ourwork,amauri}]. Thus, an attenuation of interband transitions would increase the plasma frequency, while it would also increase the critical temperature. This increase in the plasma frequency can, for example, be obtained for a lighter upper band when $\alpha_2 > 0$ in Eq.~\eqref{eq:h0}~[\cite{ourwork,amauri}]. Finally, we observe in Fig.~\ref{fig:sensitivity}(a) that the critical temperature of a single quadratic band strongly decreases for an increase in the short-range (large $q$) static screening. This effect is suppressed in a Luttinger semimetal, in Fig.~\ref{fig:sensitivity}(b), because of the weakening in the static repulsion due to the spin-orbit form-factor in the averaged Coulomb potential in~Eq.~\eqref{eq:iav}.

The superconductivity mediated by the screened Coulomb repulsion is thus mostly sensitive to plasmons and to the discontinuity of the dielectric function at $2k_F$. The plasmon mechanism occurs for larger values of $r_s$ and because the plasma frequency of a single quadratic band is larger than a Luttinger semimetal [\cite{ourwork}], it also has a larger critical temperature for larger Wigner-Seitz radii. In Luttinger semimetals, the spin-orbit correction~\eqref{eq:projector} plays a non-negligible role because it is responsible for the interband coupling that competes with the plasmon mechanism while it also weakens the short-range repulsion. Our observations for the superconductivity of Luttinger semimetals could translate to Dirac bandstructures because of the strong interband coupling [\cite{diracwp}] but which is neglected in Ref.~[\cite{ruhmanbismuth}]. 

\section{Discussion}
\label{sec:discussion}
We can analyse the applicability of our description to the superconductivity of some candidate Luttinger semimetals such as the bismuth-based half-Heuslers YPtBi, YPdBi, LuPtBi and LuPdBi [\cite{halfheuslerbs1,halfheuslerbs2,tafti,wang2013,nakajima}]. The critical temperature of these materials is in the range $T_c = 0.7 - 1.5$~K for a carrier density $n \approx 10^{19}{\rm cm}^{-3}$, a band mass $m/m_e \approx 0.1-0.3$ and a background permittivity that can be roughly estimated to $\epsilon^{*} \approx 20$ [\cite{yptbi1,yptbi2,yptbi3,yptbi4}]. This corresponds to $r_s \approx 0.5-1$ and $T_c/T_F \approx 2-8\times 10^{-4}$ which is within the order of magnitude of our calculations, where $T_c/T_F \approx 4.4(4)\times10^{-4}$. Thus, within the present mechanism of superconductivity, we expect that the gap function of these materials is a singlet $s-$wave order parameter. This stands in contrast with the recent proposition that YPtBi is a line-node superconductor from indirect evidences like the behavior of its magnetic susceptibility with temperature [\cite{linenode,splinenode}]. However, this interpretation of the measurements is arguable due to the small value of the lower critical field $B_{c1}$ [\cite{yptbimagnetic}]. Moreover, YPtBi shows deviations of its upper critical field $B_{c2}(T)$ with temperature [\cite{whh,pwhh,yptbi2}] which are not explained with the assumption of nodal superconductivity~[\cite{yptbi2}]. These discrepancies may come from the approximation of a contact pairing potential used to compute $B_{c2}(T)$ [\cite{whh,pwhh,whhlong}] which is questionable for the Coulomb potential and call for further theoretical investigation.

There is also evidence that the pyrochlore iridate Pr$_2$Ir$_2$O$_7$~[\cite{qbtir,cheng2017}] is a Luttinger semimetal with a carrier density $n \approx 10^{18}~{\rm cm}^{-3}$, a band mass $m/m_e = 6.3$ and a background dielectric constant $\epsilon^* \approx 10$, such that $r_s \approx 10 - 15$ and $T_F \approx 8$~meV. This material was studied down to $30$~mK~[\cite{smallt}] without any report of a superconducting behavior. Our model suggests that this is due to the very small Fermi temperature of this material. Using our result, $T_c/T_F \approx 4.4\times 10^{-4}$, we propose that it would be superconducting below $T_c \approx 40$~mK. The critical temperature can be lower because superconductivity would presumably compete against magnetic interactions in Pr$_2$Ir$_2$O$_7$~[\cite{smallt}].

These comparisons to experiments should however be treated with caution. First, one can question the validity of the Luttinger model for small and for large doping. Indeed, for a smaller carrier density (large $r_s$) the Coulomb interaction may lead to a non-Fermi liquid behaviour~[\cite{nfl1,nfl2,nfl3}] and to an interaction-driven topological insulator~[\cite{qbtti1}]. However, this regime with small carrier density appears difficult to observe experimentally, even in Luttinger semimetals with small Fermi temperatures such as the pyrochlore iridate Pr$_2$Ir$_2$O$_7$~[\cite{cheng2017}]. At large doping (small $r_s$), the validity of the $k\cdot P$ Hamiltonian \eqref{eq:h0} is questionable because other bands might be involved. And thus, even if we find $T_c/T_F \approx 4.4\times 10^{-4}$ down to $r_s \approx 0.01$, we expect strong deviations from this relation for large values of $T_F$.

A second reason for caution is that in the present work we only partially consider the coupling of electrons to phonons~[\cite{bcs,rickayzen,takadawphon}]. The competition of the electron-phonon coupling and the electron-electron repulsion is a long-standing issue where the Coulomb potential is usually evaluated as a perturbation~[\cite{morel,macmillan}]. In YPtBi, superconductivity due to the electron-phonon coupling~[\cite{meinert}] and the polar-phonon mechanism~[\cite{savary}] would happen for a critical temperature $T_c < 10^{-3}$~K which is much smaller than in the present theory and in the experiments, where $T_c \approx 0.7-0.9$~K [\cite{yptbi1,yptbi2,yptbi3,yptbi4}]. This suggests that the electron-phonon coupling only affects the critical temperature perturbatively in YPtBi and we thus expect no isotopic effect for this material. However, this observation cannot be generalized to all Luttinger semimetals and further work is needed to understand the situation where the electron-phonon coupling and the Coulomb repulsion compete [\cite{pseudpot,pseudpotexp}].

Another limitation of the present description is that we neglect local-field corrections to the Coulomb potential, as described by vertex corrections [\cite{singwi5,tw1,tw2,ko}]. Here, the amplitude of such terms cannot be simply related to  the ratio of some characteristic frequency to the Fermi energy, as in Migdal's theorem~[\cite{migdal}]. This was discussed in the context of superconductivity from Coulomb repulsion in a single quadratic band in~[\cite{rsvertex,rsvertexerr,brvertex,takadavertex1,takadavertex2}], where vertex corrections renormalize the critical temperatures for intermediate values of the Wigner-Seitz radius~[\cite{takadavertex2}]. Similar behaviour may also happen for Luttinger semimetals but an explicit expression of the vertex corrections is currently missing.

\section{Conclusion}
We have investigated the superconductivity of the three-dimensional quadratic band touching Luttinger semimetal from the screened Coulomb repulsion. We have derived a symmetric form of the gap equation at the critical temperature and solved it numerically. The critical temperature is linear with the Fermi temperature, $T_c/T_F \approx 4.4(4)\times 10^{-4}$, and extends to small values of the Wigner-Seitz radius, which is not the case for a single quadratic band. We used a variational principle of the gap equation to compute the sensitivity of the critical temperature to changes in the dielectric function $\epsilon(i\Omega,q)$. It shows the importance of plasmons and the discontinuity of the dielectric function at $2k_F$ in this mechanism of superconductivity, for both the single quadratic band and the quadratic band touching. The critical temperature we find is in the order of magnitude of some superconducting Luttinger semimetals, like YPtBi. Finally, we use our results to propose that the pyrochlore iridate Pr$_2$Ir$_2$O$_7$ may be superconducting below $T_c \approx 40$ mK. 

There are multiple extensions to this work, such as describing the influence of the electron-phonon pairing on the critical temperature, determining the effect of asymmetric electron and hole masses [\cite{amauri}] or introducing vertex corrections in the dielectric function. One could also study how the $s-$wave gap function competes with the other, anisotropic, superconducting order parameters proposed for Luttinger semimetals~[\cite{savary,bitan,halfheuslerbs2}]. In the context of the mechanism considered in this work, the structure of the Eliashberg equation for superconducting order parameters beyond $s-$wave was recently discussed~[\cite{qts}]. It was found that spin-orbit coupling could lead to an enhancement in the $\ell = 1$ channel for instance. A numerical calculation will be needed to identify the preferred channel. Finally, because the magnetic response of superconductors is usually computed by assuming a contact pairing potential~[\cite{whh,pwhh,whhlong}], it is worth considering how accurately it applies to pairing from the Coulomb repulsion.\\

\begin{acknowledgments}
We would like to thank E. Dupuis, M. Comin and V. Kaladzhyan for fruitful discussions. This project is funded by a grant from Fondation Courtois, a Discovery Grant from NSERC, a Canada Research Chair, and a ``\'Etablissement de nouveaux chercheurs et de nouvelles chercheuses universitaires'' grant from FRQNT. This research was enabled in part by support provided by Calcul Qu\'ebec (www.calculquebec.ca) and Compute Canada (www.computecanada.ca)
\end{acknowledgments} 

\bibliographystyle{apsrev4-1}
\bibliography{bibliography}

\appendix 

\begin{widetext}
\section{Eliashberg equation for singlet superconductivity}
\label{app:elshberg}

In this section we derive the Eliashberg equation due to the electron-electron repulsion with account of screening, self-energy corrections and for a pseudo-spin singlet pairing, \emph{i.e.} with opposite Kramer partners within a band. Because of rotation symmetry, the Eliashberg equation can be decomposed on the spherical harmonics $Y_{\ell,m}(\theta,\phi)$ and in main text we only discuss the situation where $\ell = 0$ ($s-$wave channel).

The eigenstates $| \sigma, \lambda, {\bf k} \rangle$ of the Hamiltonian $\hat{H}_0$ \eqref{eq:h0} define the fermionic operators $\hat{c}_{{\bf k}\sigma\lambda} = \sum_{s} \langle \sigma, \lambda, {\bf k} | \psi_{{\bf k}s} \rangle \hat{\psi}_{{\bf k}s}$, where $\sigma = \pm$ indicates the upper or lower subband, $\lambda = \pm$ the Kramer partners within a subband and $s = \{3/2,1/2,-1/2,-3/2\}$ indicates the eigenvalues of $\hat{J}_z$ for the $j = 3/2$ fermions. In this basis, the normal-ordered Hamiltonian is
\begin{align}
    \hat{H} = \sum_{{\bf k}, \sigma\lambda}& \xi_{\sigma}({\bf k}) \hat{c}^{\dagger}_{{\bf k},\sigma\lambda} \hat{c}_{{\bf k},\sigma\lambda}\\
    &+ \frac{1}{2\mathcal{V}} \sum_{{\bf q} \neq 0}\sum_{\substack{ {\bf k}_1\sigma_1\tau_1\\\sigma_3\lambda_3}}\sum_{\substack{ {\bf k}_2\sigma_2\tau_2\\\sigma_4\lambda_4}} V_0(q) \langle \sigma_3\lambda_3 {\bf k}_1 + {\bf q}|\sigma_1\lambda_1 {\bf k}_1\rangle \langle \sigma_4\lambda_4 {\bf k}_2 - {\bf q}|\sigma_2\lambda_2 {\bf k}_2 \rangle\hat{c}^{\dagger}_{{\bf k}_1+{\bf q}\sigma_3\lambda_3} \hat{c}^{\dagger}_{{\bf k}_2-{\bf q}\sigma_4\lambda_4} \hat{c}_{{\bf k}_2\sigma_2\lambda_2}\hat{c}_{{\bf k}_1\sigma_1\lambda_1}.\nonumber
\end{align}
We consider the equations of motion of the Green's functions $G_{\sigma\lambda}(\tau, {\bf p}) = \left\langle \hat{T}_{\tau} \hat{c}_{{\bf p}\sigma\lambda}(\tau) \hat{c}_{{\bf p}\sigma\lambda}^{\dagger}(0) \right\rangle$, $F_{\sigma}(\tau, {\bf p}) = \left\langle \hat{T}_{\tau} \hat{c}_{{\bf p}\sigma +}(\tau) \hat{c}_{-{\bf p}\sigma -}(0) \right\rangle$ and $F^{*}_{\sigma}(\tau, {\bf p}) = \left\langle \hat{T}_{\tau} \hat{c}^{\dagger}_{-{\bf p}\sigma -}(\tau) \hat{c}^{\dagger}_{{\bf p}\sigma +}(0) \right\rangle$, for singlet superconductivity
\begin{align}
    \left( \frac{\partial}{\partial \tau} + \xi_{\sigma}({\bf p}) \right) &G_{\sigma\lambda}(\tau, {\bf p}) 
    = \delta(\tau)\\
    &- \sum_{{\bf q} \neq 0}V_{0}({ q}) \sum_{\substack{{\bf k} \sigma_1\sigma_2 \sigma^{\prime}\\ \lambda_1\lambda_2 \lambda^{\prime}}} \langle \sigma \lambda {\bf p}|\sigma^{\prime} \lambda^{\prime} {\bf p + q }\rangle \langle \sigma_1\lambda_1 {\bf k }|\sigma_2\lambda_2 {\bf k - q}\rangle \left\langle \hat{T}_{\tau} \hat{c}^{\dagger}_{{\bf k}\sigma_1\lambda_1}(\tau) \hat{c}_{{\bf k}-{\bf q}\sigma_2\lambda_2}(\tau) \hat{c}_{{\bf p}+{\bf q}\sigma^{\prime} \lambda^{\prime}}(\tau) \hat{c}^{\dagger}_{{\bf p}\sigma\lambda}(0) \right\rangle,\nonumber
    \\
    \left( \frac{\partial}{\partial \tau} + \xi_{\sigma}({\bf p}) \right) &F_{\sigma}(\tau, {\bf p}) 
    =\\
    &- \sum_{{\bf q} \neq 0} V_{0}({ q}) \sum_{\substack{{\bf k} \sigma_1\sigma_2 \sigma^{\prime}\\ \lambda_1\lambda_2 \lambda^{\prime}}} \langle \sigma +, {\bf p}|\sigma^{\prime} \lambda^{\prime} {\bf p + q }\rangle \langle \sigma_1\lambda_1 {\bf k }|\sigma_2\lambda_2 {\bf k - q}\rangle \left\langle \hat{T}_{\tau} \hat{c}^{\dagger}_{{\bf k}\sigma_1\lambda_1}(\tau) \hat{c}_{{\bf k}-{\bf q}\sigma_2\lambda_2}(\tau) \hat{c}_{{\bf p}+{\bf q}\sigma^{\prime} \lambda^{\prime}}(\tau) \hat{c}_{-{\bf p}\sigma -}(0) \right\rangle\nonumber.
\end{align}
The retardation effects are included by deriving the time evolution of $\left\langle \hat{T}_{\tau} \hat{c}^{\dagger}_{{\bf k}\sigma_1\lambda_1}(\tau) \hat{c}_{{\bf k}-{\bf q}\sigma_2\lambda_2}(\tau) \hat{O} \right\rangle$ in the random phase approximation (RPA). This is similar to the retardation effects from the electron-phonon coupling [\cite{rickayzen}]. The detailed calculation can be found in Appendix~\ref{app:rpa} and leads to
\begin{align}
    V_{0}({ q})\sum_{\substack{{\bf k}\sigma_1\sigma_2\\\lambda_1\lambda_1}} \langle \sigma_1\lambda_1 {\bf k}&|\sigma_2\lambda_2 {\bf k} - {\bf q} \rangle\left\langle \hat{T}_{\tau} \hat{c}^{\dagger}_{{\bf k}\sigma_1\lambda_1}(\tau) \hat{c}_{{\bf k}-{\bf q}\sigma_2\lambda_2}(\tau) \hat{O} \right\rangle \\
    &= \int d\tau^{\prime} V(\tau - \tau^{\prime},{\bf q}) \sum_{\substack{{\bf k}\sigma_1\sigma_2\\\lambda_1\lambda_1}} \langle \sigma_1\lambda_1 {\bf k}|\sigma_2\lambda_2 {\bf k} - {\bf q} \rangle\left\langle \hat{T}_{\tau} \hat{c}^{\dagger}_{{\bf k}\sigma_1\lambda_1}(\tau^{\prime}) \hat{c}_{{\bf k}-{\bf q}\sigma_2\lambda_2}(\tau^{\prime}) \hat{O} \right\rangle,\nonumber
\end{align}
where $V(\tau,{\bf q})$ is the screened Coulomb potential. After including this retardation effect in the four-operators Green's functions, we decompose them over the normal and anomalous Green's functions with Wick's decomposition. For example,
\begin{align}
    \sum_{\substack{{\bf k} \sigma_1\sigma_2 \sigma^{\prime}\\ \lambda_1\lambda_2 \lambda^{\prime}}}\langle \sigma \lambda {\bf p}|\sigma^{\prime} \lambda^{\prime} {\bf p + q }\rangle &\langle \sigma_1\lambda_1 {\bf k }|\sigma_2\lambda_2 {\bf k - q}\rangle \left\langle \hat{T}_{\tau} \hat{c}^{\dagger}_{{\bf k}\sigma_1\lambda_1}(\tau^{\prime}) \hat{c}_{{\bf k}-{\bf q}\sigma_2\lambda_2}(\tau^{\prime}) \hat{c}_{{\bf p}+{\bf q}\sigma^{\prime} \lambda^{\prime}}(\tau) \hat{c}^{\dagger}_{{\bf p}\sigma\lambda}(0) \right\rangle\\
    &= \left( \sum_{\sigma^{\prime}\lambda^{\prime}} |\langle \sigma \lambda{\bf p}|\sigma^{\prime}\lambda^{\prime} {\bf p + q }\rangle|^2 \left\langle \hat{T}_{\tau} \hat{c}_{{\bf p}+{\bf q}\sigma^{\prime}\lambda^{\prime}}(\tau) \hat{c}^{\dagger}_{{\bf p}+{\bf q}\sigma^{\prime}\lambda^{\prime}}(\tau^{\prime}) \right\rangle \right) \left\langle \hat{T}_{\tau} \hat{c}_{{\bf p}\sigma\lambda}(\tau^{\prime}) \hat{c}^{\dagger}_{{\bf p}\sigma\lambda}(0) \right\rangle\\
    &- \sum_{\sigma^{\prime} \lambda^{\prime}} \langle \sigma\lambda {\bf p}|\sigma^{\prime} \lambda^{\prime} {\bf p + q }\rangle \langle \sigma\bar{\lambda}, -{\bf p }|\sigma^{\prime} \bar{\lambda}^{\prime} {\bf -p - q}\rangle \left\langle \hat{T}_{\tau} \hat{c}_{-{\bf p}-{\bf q}\sigma^{\prime} \bar{\lambda}^{\prime}}(\tau^{\prime}) \hat{c}_{{\bf p}+{\bf q}\sigma^{\prime} \lambda^{\prime}}(\tau) \right\rangle \left\langle \hat{T}_{\tau} \hat{c}^{\dagger}_{{\bf p}\sigma\lambda}(0) \hat{c}^{\dagger}_{-{\bf p} \sigma\bar{\lambda} }(\tau^{\prime}) \right\rangle\nonumber.
\end{align}
where $\bar{\lambda} = -\lambda$. The index $\lambda = \pm$ describes Kramer partners so, because the system is time-reversal symmetric, one has $\langle \sigma\lambda {\bf p}|\sigma^{\prime} \lambda^{\prime} {\bf p + q }\rangle \langle \sigma\bar{\lambda}, -{\bf p }|\sigma^{\prime} \bar{\lambda}^{\prime} {\bf -p - q}\rangle = |\langle \sigma\lambda {\bf p}|\sigma^{\prime} \lambda^{\prime} {\bf p + q }\rangle|^2 $ and
\begin{align}
    \label{eq:app:tsymF}
    \sum_{\substack{{\bf k} \sigma_1\sigma_2 \sigma^{\prime}\\ \lambda_1\lambda_2 \lambda^{\prime}}}&\langle \sigma \lambda {\bf p}|\sigma^{\prime} \lambda^{\prime} {\bf p + q }\rangle \langle \sigma_1\lambda_1 {\bf k }|\sigma_2\lambda_2 {\bf k - q}\rangle \left\langle \hat{T}_{\tau} \hat{c}^{\dagger}_{{\bf k}\sigma_1\lambda_1}(\tau^{\prime}) \hat{c}_{{\bf k}-{\bf q}\sigma_2\lambda_2}(\tau^{\prime}) \hat{c}_{{\bf p}+{\bf q}\sigma^{\prime} \lambda^{\prime}}(\tau) \hat{c}^{\dagger}_{{\bf p}\sigma\lambda}(0) \right\rangle\\
    =& \left(\sum_{\sigma^{\prime}\lambda^{\prime}} |\langle \sigma \lambda{\bf p}|\sigma^{\prime}\lambda^{\prime} {\bf p + q }\rangle|^2 G_{\sigma^{\prime} \lambda^{\prime}}(\tau-\tau^{\prime},{\bf p}+{\bf q}) \right)G_{\sigma\lambda}(\tau^{\prime},{\bf p}) - \sum_{\sigma^{\prime} \lambda^{\prime}} |\langle \sigma\lambda {\bf p}|\sigma^{\prime} \lambda^{\prime} {\bf p + q }\rangle|^2 F_{\sigma^{\prime}}(\lambda^{\prime}(\tau-\tau^{\prime}),\lambda^{\prime}({\bf p}+{\bf q}))F^{*}_{\sigma}(\lambda \tau^{\prime},\lambda{\bf p}).\nonumber
\end{align}
The equations being time-reversal and inversion symmetric, one can consider different anomalous Green's function with even or odd parity in either time and wavevector
\begin{align}
    \label{eq:app:sympar}
    F_{\sigma}(\tau,{\bf p}) = (-1)^{\pi_T} F_{\sigma}(-\tau,{\bf p}) = (-1)^{\pi_T + \pi_I} F_{\sigma}(-\tau,-{\bf p}),
\end{align}
with $\pi_T, \pi_I \in \{0,1\}$. Then the anomalous Green's functions in the last line of Eq.~\eqref{eq:app:tsymF} are
\begin{align}
    \label{eq:app:symeq}
    F_{\sigma^{\prime}}(\lambda^{\prime}(\tau-\tau^{\prime}),\lambda^{\prime}({\bf p}+{\bf q})) = \lambda^{\prime \pi_T + \pi_I}F_{\sigma^{\prime}}(\tau-\tau^{\prime},{\bf p}+{\bf q})
\end{align}
and the summation identically vanishes if $\pi_T + \pi_I$ is odd. A non-zero gap function must be either even or odd for \emph{both} time and space, and in the following we only consider the situation of an even gap function (\emph{i.e.} $\pi_T = \pi_I = 0$).

We replace these Wick decompositions in the original equation which we also Fourier transform over Matsubara frequencies $\omega_n = (2n+1)\pi T$ 
\begin{align}
    G_{\sigma\lambda}(\tau_1-\tau_2,{\bf p}) &= T \sum_{\omega_n} G_{\sigma\lambda}(i\omega_{n},{\bf p}) e^{-i\omega_{n}(\tau_1-\tau_2)},\\
    F_{\sigma}(\tau_1-\tau_2,{\bf p}) &= T \sum_{\omega_n} F_{\sigma}(i\omega_{n},{\bf p}) e^{-i\omega_{n}(\tau_1-\tau_2)}.
\end{align}
This leads to the following set of equations
\begin{align}
    &(-i\omega_n + \xi_{\sigma}({\bf p}))G_{\sigma\lambda}(i\omega_{n},{\bf p}) = 1 - \Sigma_{\sigma\lambda}(i\omega_{n},{\bf p}) G_{\sigma\lambda}(i\omega_{n},{\bf p}) - \Delta_{\sigma}(i\omega_{n},{\bf p})F^{*}_{\sigma}(i\omega_{n},{\bf p}),\\
    &(-i\omega_n + \xi_{\sigma}({\bf p}))F_{\sigma}(i\omega_{n},{\bf p}) = - \Sigma_{\sigma +}(i\omega_{n},{\bf p}) F_{\sigma}(i\omega_{n},{\bf p}) + \Delta_{\sigma}(i\omega_{n},{\bf p})G_{\sigma -}(-i\omega_{n},-{\bf p}),\\
    &(-i\omega_n - \xi_{\sigma}(-{\bf p}))F^{*}_{\sigma}(i\omega_{n},{\bf p}) = \Sigma_{\sigma -}(-i\omega_{n},-{\bf p}) F^{*}_{\sigma}(i\omega_{n},{\bf p}) - \bar{\Delta}_{\sigma}(i\omega_{n},{\bf p})G_{\sigma +}(i\omega_{n},{\bf p}),
\end{align}
where we have introduced the normal and anomalous self-energies
\begin{align}
    \Sigma_{\sigma\lambda}(i\omega_{n},{\bf p}) &= -T\sum_{{\bf q} \neq 0 \omega_m \sigma^{\prime}\lambda^{\prime}} V(i(\omega_n - \omega_m),{\bf q}) |\langle \sigma\lambda {\bf p}|\sigma^{\prime} \lambda^{\prime} {\bf p + q }\rangle|^2 G_{\sigma^{\prime} \lambda^{\prime}}(i\omega_{m},{\bf p}+{\bf q}),\\
    \Delta_{\sigma}(i\omega_{n},{\bf p}) &=  -T\sum_{{\bf q} \neq 0 \omega_m \sigma^{\prime}\lambda^{\prime}} V(i(\omega_n - \omega_m),{\bf q}) |\langle \sigma\lambda {\bf p}|\sigma^{\prime} \lambda^{\prime} {\bf p + q }\rangle|^2 F_{\sigma^{\prime}}(i\omega_{m}, {\bf p}+{\bf q}),\nonumber\\
    \bar{\Delta}_{\sigma}(i\omega_{n},{\bf p}) &= -T \sum_{{\bf q} \neq 0 \omega_m \sigma^{\prime}\lambda^{\prime}} V(i(\omega_n - \omega_m),{\bf q}) |\langle \sigma\lambda {\bf p}|\sigma^{\prime} \lambda^{\prime} {\bf p + q }\rangle|^2 F^{*}_{\sigma^{\prime}}(i\omega_{m},{\bf p}+{\bf q}),\nonumber
\end{align}
with $V(i\Omega_n,{\bf q}) = V_{0}({q})/\epsilon(i\Omega_n,{\bf q})$ the screened Coulomb potential.

In the following we write the expression for the normal and anomalous self-energy that we consider throughout our work. There, we study the phase transition from the normal to the superconducting phase, that is the temperature $T = T_c$ beyond which the anomalous Green's function $F_{\sigma}$ vanishes ($F_{\sigma} = 0$). 

\subsection{Self-energy in the normal phase}
\label{sec:app:selfe}

In the normal state ($F_{\sigma} = 0$) the normal Green's functions are
\begin{align}
    G^{(N)}_{\sigma\lambda}(i\omega_n,{\bf p}) = \left( -i\omega_n + \xi_{\sigma}({\bf p}) + \Sigma_{\sigma\lambda}(i\omega_n, {\bf p}) \right)^{-1},
\end{align}
where the self-energies in the normal phase are
\begin{align}
    \label{eq:app:self}
    \Sigma^{(N)}_{\sigma\lambda}(i\omega_n,{\bf p}) &= -\beta^{-1} \sum_{\Omega_m \sigma^{\prime}\lambda^{\prime}{\bf q} } \frac{V_0({ q})}{\epsilon(i\Omega_m,{\bf q})} |\langle \sigma\lambda {\bf p}|\sigma^{\prime}\lambda^{\prime} {\bf p + q }\rangle|^2 G^{(N)}_{\sigma^{\prime}\lambda^{\prime}}(i(\omega_n - \Omega_{m}),{\bf p}+{\bf q})\\
        &\approx - ({2\beta})^{-1} \sum_{\Omega_m \sigma^{\prime} {\bf q} } \frac{V_0({ q})}{\epsilon(i\Omega_m,{\bf q})}{\rm Tr}\left( \hat{P}_{\sigma}({\bf p}) \hat{P}_{\sigma^{\prime}}({\bf p}+{\bf q}) \right) G^{(N0)}_{\sigma^{\prime}}(i(\omega_n - \Omega_{m}),{\bf p}+{\bf q}),
\end{align}
with the bosonic Matsubara frequencies $\Omega_m = 2m\pi T$ with $T$ the temperature and $m$ an integer. In the last line we have approximated $G^{(N)}_{\sigma\lambda}(i\omega_n,{\bf p}) \approx G^{(N0)}_{\sigma}(i\omega_n,{\bf p}) = (-i\omega_n + \xi_{\sigma}({\bf p}))^{-1}$ which is independent of the index $\lambda$. We have discussed this expression of the self-energy for real frequencies in Ref.~[\cite{ourwork}] and with the same approach we compute its behaviour for imaginary frequencies in Fig.~\ref{fig:spenrg}(b,c). There we decompose the self-energy over two real-valued functions $Z_{\sigma}(i \omega_n,{\bf p})$ and $\chi_{\sigma}(i\omega_n,{\bf p})$
\begin{align}
    \Sigma_{\sigma}(i\omega_n,{\bf p}) = \chi_{\sigma}(i\omega_n,{\bf p}) + i \omega_n(1 - Z_{\sigma}(i \omega_n,{\bf p})).
\end{align}

\subsection{Anomalous self-energy}
\label{sec:app:aself}

The anomalous Green's functions satisfy
\begin{align}
    \label{eq:appF}
    F_{\sigma}(i\omega_n, {\bf p}) = \left( -i\omega_n + \xi_{\sigma}({\bf p}) + \Sigma_{\sigma +}(i\omega_n, {\bf p}) \right)^{-1} G_{\sigma -}(-i\omega_n, -{\bf p}) \Delta_{\sigma}(i\omega_n,{\bf p})
\end{align}
where the anomalous self-energy is
\begin{align}
    \Delta_{\sigma}(i\omega_n,{\bf p}) &=  - T\sum_{\sigma^{\prime}\lambda^{\prime}\omega_m,\bf q \neq 0} V(i(\omega_n - \omega_m),{\bf q}) |\langle \sigma\lambda {\bf p}|\sigma^{\prime} \lambda^{\prime} {\bf p + q }\rangle|^2 F_{\sigma^{\prime}}(i\omega_{m}, {\bf p}+{\bf q}).
\end{align}
Close to the critical temperature one can neglect the amplitude of the anomalous Green's functions, $F_{\sigma} \approx 0$. The normal Green's functions can also be approximated with their normal state behaviour that we have discussed in Ref.~[\cite{ourwork}], $G_{\sigma \lambda}(i\omega_n, {\bf p}) \approx G^{(N)}_{\sigma}(i\omega_n, {\bf p})$. This leads to the linearised Eliashberg equations that we discuss in the next subsections and that we transform to have it symmetric.

\subsubsection{Linear Eliashberg equations}

Near the phase transition, $T = T_c$, we expand Eq.~\eqref{eq:appF} to the lowest order in $F_{\sigma}$ 
\begin{align}
    F_{\sigma_1}(i\omega_{n_1},{\bf k}_1) = - G_{\sigma_1}(i\omega_{n_1},{\bf k}_1)G_{\sigma_1}(-i\omega_{n_1},-{\bf k}_1)T\sum_{\omega_{n_2}}\sum_{\sigma_2 \bf k_2} I_{\sigma_1\sigma_2}(i\omega_{n_1},{\bf k}_1; i\omega_{n_2}, {\bf k}_2) F_{\sigma_2}(i\omega_{n_2},{\bf k}_2)
\end{align}
where the coupling between electrons in a pair is described by
\begin{align}
    I_{\sigma_1\sigma_2}(i\omega_{n_1},{\bf k}_1; i\omega_{n_2}, {\bf k}_2) = \frac{V_0(|{\bf k}_1 - {\bf k}_2|)}{\epsilon(i(\omega_{n_1}-\omega_{n_2}), {\bf k}_1 - {\bf k}_2)} \frac12 {\rm Tr}\left( \hat{P}_{\sigma_1}({\bf k}_1) \hat{P}_{\sigma_2}({\bf k}_2) \right).
    \label{eq:app:iproj}
\end{align}
This equation is similar to that developed in previous studies on the superconductivity mediated by plasmons [\cite{takada}] up to the spin-orbit form factor $\frac12 {\rm Tr}\left( \hat{P}_{\sigma_1}({\bf k}_1) \hat{P}_{\sigma_2}({\bf k}_2) \right)$.

We introduce the gap functions $\phi_{\sigma}(i\omega_n,{\bf k}) = G^{-1}_{\sigma}(i\omega_{n},{\bf k})G^{-1}_{\sigma}(-i\omega_{n},-{\bf k}) F_{\sigma}(i\omega_n, {\bf k})$ for which the linearised Eliashberg equation writes
\begin{align}
    \phi_{\sigma_1}(i\omega_{n_1},{\bf k}_1) = - T \sum_{\omega_{n_2}}\sum_{\sigma_2 {\bf k}_2} \frac{I_{\sigma_1\sigma_2}(i\omega_{n_1},{\bf k}_1; i\omega_{n_2}, {\bf k}_2)}{(\omega_{n_2} Z_{\sigma_2}(i\omega_{n_2},{\bf k}_2))^2 + (\xi_{\sigma_2}({\bf k}_2) + \chi_{\sigma_2}(i\omega_{n_2},{\bf k}_2))^2} \phi_{\sigma_2}(i\omega_{n_2},{\bf k}_2).
\end{align}
Due to the rotational symmetry of the non-interacting Hamiltonian and of the Coulomb interaction, the electron-electron coupling $I_{\sigma_1\sigma_2}(i\omega_{n_1},{\bf k}_1, i\omega_{n_2}, {\bf k}_2)$ depends on ${\bf k}_1$ and ${\bf k}_2$ through their norms $k_1 = |{\bf k}_1|$, $k_2 = |{\bf k}_2|$ and their relative angle $\theta_{{\bf k}_1,{\bf k}_2}$. This allows to decompose the gap functions over the spherical harmonics $Y_{\ell m}(\theta,\phi)$ 
\begin{align}
    \phi_{\sigma}(i\omega_n,{\bf k}) = \sum_{\ell m} \phi_{\ell m\sigma}(i\omega_n,k) Y_{\ell m}(\theta,\phi)
\end{align}
where $\theta$ is the angle between ${\bf k}$ and the $z-$axis and $\phi$ is the angle between the projection of ${\bf k}$ in the $xy$ plane and the $x-$axis. The Eliashberg equation is degenerate on the index $m$ and the equations for the components $\phi_{\ell}$ are
\begin{align}
    \label{eq:appelshphiz}
    \phi_{ \ell  \sigma_1}(i\omega_{n_1},k_1) = -T&\sum_{\sigma_2\omega_{n_2}} \int_{0}^{\infty} dk_2~ \frac{k_2}{k_1} \frac{I_{\ell\sigma_1\sigma_2}(i\omega_{n_1}, k_1 ; i\omega_{n_2}, k_2)}{(\omega_{n_2} Z_{\sigma_2}(i\omega_{n_2},{ q}))^2 + \left(\xi_{\sigma_2}({ q}) + \chi_{\sigma_2}(i\omega_{n_2},{ q})\right)^2}\phi_{\ell\sigma_2 }(i\omega_{n_2},k_2)
\intertext{with}
    I_{\ell\sigma_1\sigma_2}(i\omega_{n_1}, k_1 ; i\omega_{n_2}, k_2) = &\int_{|k_1-k_2|}^{k_1+k_2} dq~P_{\ell}\left(\frac{k_1^2 +k_2^2 - q^2}{2k_1k_2}\right) \frac14\left\{ 2 + \sigma_1 \sigma_2 \left[ 3 \left(\frac{k_1^2 +k_2^2 - q^2}{2k_1k_2}\right)^2 - 1 \right] \right\} \frac{q V_0(q) N_0}{\epsilon(i(\omega_{n_1} - \omega_{n_2}),q)},
\end{align}
and where $N_0 = 1/(4\pi^2)$ is the density of states per band at the Fermi surface. This equation is similar to that derived by Takada in Ref.~[\cite{takada}] and accounts for the spin-orbit corrections of a Luttinger semimetal, as in Ref.~[\cite{savary}]. In practice we consider a gap function even in frequency, following our comment after Eq.~\eqref{eq:app:tsymF}, and we can symmetrize the equation to have the summation over positive Matsubara frequencies. This choice necessitates that the gap functions are even in momentum, which excludes odd spherical harmonics. We observe that in Ref.~[\cite{takada}] the author finds solutions for a singlet gap function, even in frequency and with $\ell = 1$ ($p-$wave) but it seems that the parity considerations in Eq.~\eqref{eq:app:sympar} were omitted in his derivation of the Eliashberg equation. 

The numerical treatment of this equation was discussed in Ref.~[\cite{takada}] where the gap function is made discrete over $i\omega_n$ and $k$, such that for $\omega_s < \omega_n < \omega_{s+1}$ and $k_d < k < k_{d+1}$, $\phi_{\sigma\ell}(i\omega_n, k) = \Delta_{\sigma\ell}(s,d)$. The equation then resembles to an eigenvalue equation
\begin{align}
    \lambda(T) { \Delta}_{\ell} = \hat{{M}}_{\ell} \cdot { \Delta}_{\ell},
\end{align}
for which we determine $T_c$ by searching for the temperature where $\lambda(T) = 1$. 

\subsubsection{Symmetrized linear Eliashberg equations}

In the main text we used a different formulation of Eq.~\eqref{eq:appelshphiz}. We have performed the transformation 
\begin{align}
    \bar{\phi}_{\ell\sigma}(i\omega_n,{k}) = \frac{k}{({\omega}_n Z_{\sigma}(i\omega_n,k))^2 + (\xi_{\sigma}({k}) + \chi_{\sigma}(i\omega_n,k))^2}\phi_{\ell\sigma}(i\omega_n,{k})
\end{align}
so that we instead have the following eigenvalue equation, with $\rho(T = T_c) = 0$,
\begin{align}
    \label{eq:appsym}
    \rho \bar{\phi}_{\ell\sigma_1}(i\omega_{n_1},k_1) = -&\sum_{\omega_{n_2}}\int_{0}^{\infty}dk_2 \bigg\{ I_{\ell\sigma_1\sigma_2}(i\omega_{n_1}, k_1 ; i\omega_{n_2}, k_2)\nonumber\\
    & + \delta_{n_1 n_2}\delta_{{ k}_1{ k}_2} \delta_{\sigma_1\sigma_2}T^{-1}\left(({\omega}_{n_2} Z_{\sigma_2}(k_2,i\omega_{n_2}))^2 + (\xi_{\sigma_2}(k_2) + \chi_{\sigma_2}(k_2,i\omega_{n_2}))^2\right) \bigg\}\bar{\phi}_{\ell\sigma_2}(i\omega_{n_2},{ k}_2).
\end{align}
The asymptotic behaviour shows that the parameter $\rho$ satisfies $\rho(T) \sim -1/T$ as $T \rightarrow 0$ and $\rho(T) \sim - T$ as $T \rightarrow \infty$. Thus, the largest possible eigenvalue $\rho(T)$ at a fixed temperature vanishes at the highest critical temperature~[\cite{bergmannrainer,allendynes}]
\begin{align}
    \rho_{\rm max}(T = T_c) = 0.
\end{align}
The equation \eqref{eq:appsym} is symmetric when permuting the indices $(\sigma_1,i\omega_{n_1},k_1) \leftrightarrow (\sigma_2,i\omega_{n_2},k_2)$. Thus, for any trial gap function $\bar{\phi}^{\rm t}$, one has the following variational principle 
\begin{align}
    \rho_{\rm max} \geq \bar{\phi}^{\rm t}\cdot \hat{S}\bar{\phi}^{\rm t}/(\bar{\phi}^{\rm t}\cdot \bar{\phi}^{\rm t}) = \rho^{\rm t},
\end{align}
where $\rho\bar{\phi}_{\ell} = \hat{S} \bar{\phi}_{\ell}$ represents Eq.~\eqref{eq:appsym}. The scalar product refers to the canonical scalar product on indices $(\sigma,i\omega_{n},k)$. This inequality implies that any critical temperature $T_c^{\rm t}$ one computes numerically with Eq.~\eqref{eq:appsym} is bounded from above by the analytic solution, $T_c > T_c^{\rm t}$. 
This formulation is helpful when computing the variational derivative of the critical temperature over $\epsilon(i\Omega_n,q)$ due to the Hellmann-Feynman theorem (see Sec.~\ref{sec:sensitivity}).

\begin{figure}
    \centering
    \includegraphics[width = \textwidth]{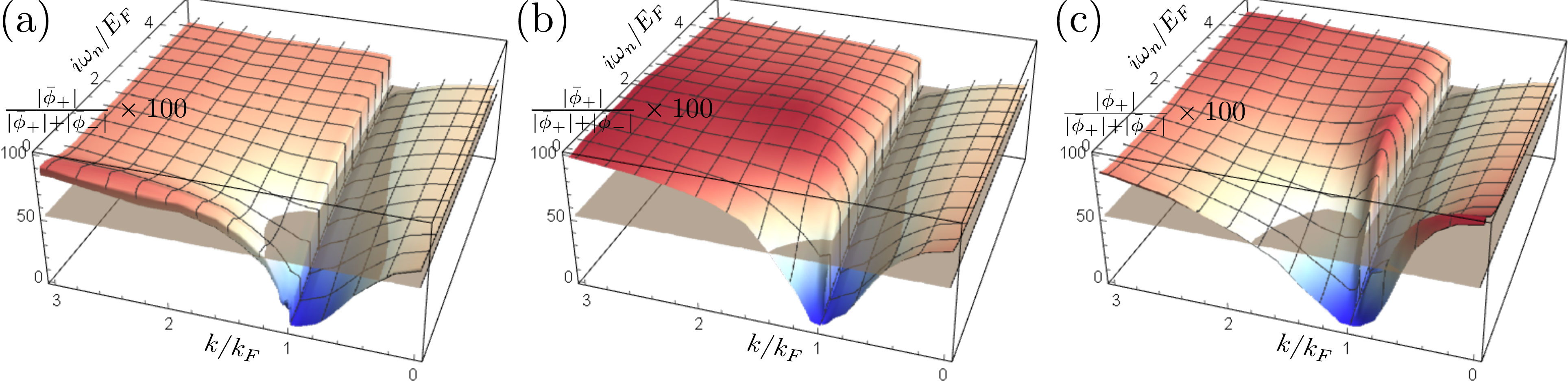}
    \caption{Proportion of $|\bar{\phi}_{+}(i\omega_n,k)|$ in percentage of $|\bar{\phi}_{-}(i\omega_n,k)| + |\bar{\phi}_{+}(i\omega_n,k)|$ for $\ell = 0$ ($s-$wave) and (a) $r_s = 1$, (b) $r_s = 5$ and (c) $r_s = 15$. The components of $\bar{\phi}_+$ that dominate over $\bar{\phi}_-$ are above the gray plane, at $50\%$, and this happens away from the Fermi surface where $k = k_F$ and $\omega_n \approx 0$.}
    \label{fig:gapratio}
\end{figure}

\subsubsection{Grid and asymptotic behaviour}
\label{app:asymptotic}

The numerical solution of the symmetrized linear Eliashberg equation is obtained by decomposing the gap function over a grid in frequencies and wavevectors (see Sec.~\ref{subsec:num}). We have refined the grid points in order to obtain a stable solution for the critical temperature. The eigenvalues $\rho(T)$ are computed using a C implementation of the LAPACK library (Intel MKL). It is worth mentioning that we are limited by the precision of the numerical variables, which are double precision, and we observe numerical errors for temperatures below $T/T_F \approx 10^{-5}$.

The two components, $\bar{\phi}_{\pm}$, of the gap functions are accounted for. We observe that $\bar{\phi}_{+}$ is non-negligible away from $k \approx k_F$ and small $\omega_n$ as depicted in Fig.~\ref{fig:gapratio}. This is because both the $s-$wave pairing potential \eqref{eq:iav} and the kinetic energy \eqref{eq:Ks} are almost independent on the band index, $\sigma$, for $k$ far away from $k_F$ and large frequency. In the case of non-$s-$wave pairing channels, which we do not consider in the present work, the contributions of the two bands may be asymmetric due to their respective helicity [\cite{savary}].

The critical temperature is small compared to the characteristic energy scale of the dielectric function ($T_c/T_F \approx 4.4\times 10^{-4}$) and we have to use a grid in frequency that can account for both scales. The numerical results reported in this work are obtained with a grid of $80$ logarithmically-spaced frequencies on the range $[10^{-6},10]$. We start with such a small frequency to be able to compute $\rho(T)$ for smaller temperatures and compute its derivative $\partial \rho/\partial T$ in Eq.~\eqref{eq:tride}. To this we add $20$ linearly-spaced frequencies up to $10^{3}$ in order to reach the expected asymptotic behaviour (see below). We do not have to introduce negative frequencies because we consider gap functions even in frequency. 

The diagonal elements in the gap equation \eqref{eq:eshsym} are dominated by the matrix elements of $K_{\sigma}$ defined in Eq.~\eqref{eq:Ks}. The smallest value of $K_{\sigma}$ is obtained for $\sigma = -$ , the $n = 0$ Matsubara frequency and $k = k_F$, where its value is $K_{-}( n = 0, k = k_F ) = Z_F^2 \pi^2 T$. Here, because we average the equation on extended intervals $[k_d,k_d+1]$, on the grid of wavevectors $\{k_d\}_{d\in[1,N_2]}$, this minimal value is only obtained for a very dense grid near $k = k_F$. Indeed, the average of $K_{-}$ \eqref{eq:Ks} on the interval centered around $k = k_F$, $[1-\Delta k/2, 1+ \Delta k/2]$ is
\begin{align}
    \frac{1}{\Delta k}\int_{1-\Delta k/2}^{1+\Delta k/2} dk K_{-}(n = 0,k) = ( \pi^2 T^2 + (1/3 + \Delta k^2/80)\Delta k^2)/T,
\end{align}
where we neglect self-energy corrections to make the expression simple. We thus properly describe the excitations at the Fermi surface for $T_c/T_F \sim 10^{-4}$ if $\Delta k/k_F \ll 10^{-4}$ near $k = k_F$. The necessity for such a narrow grid near the Fermi surface is also present in other related works [\cite{takada,ashcroft,ruhmanbismuth}] and is seen as a dip in the resulting gap function (see Fig.~\ref{fig:gap12}). We used a dense grid with $60$ points in the interval $k/k_F\in[0.99,1.01]$ with the smallest interval of $\Delta k/k_F = 10^{-6}$, which constitutes the smallest spacing we can reach here with double numerical precision. This tight spacing close to the Fermi surface limits the exploration of the critical temperature down to $T/T_F \approx 10^{-5}$. To this we add $12$ points in the interval $[0,0.98]$ and $14$ points in the interval $[1.02,5]$. This gives a smooth behaviour away from $k \approx k_F$ and allows to describe the asymptotic behaviour.

The asymptotic behaviour of $\bar{\phi}_{\ell\sigma}(i\omega_n,k)$ is independent on $\ell$ and can be determined from that of $\phi_{\ell\sigma}(i\omega_n,k)$ in Ref.~[\cite{rspseudopot}]
\begin{align}
    \bar{\phi}_{\ell\sigma}(i\omega_n,k) \sim \left\{
    \begin{array}{ll}
        1/\omega_n^2 & \textrm{for } \omega_n \gg E_F,\\
        1/k^{5} & \textrm{for } k \gg k_F.
    \end{array}
    \right.
\end{align}
We use this asymptotic behaviour to describe the large frequency and large wavevector behaviour beyond a frequency $\omega_{N_1}$ and a wavevector $k_{N_2}$ (see Sec.~\ref{subsec:num}). We check that the gap function indeed converges to these asymptotic behaviours by plotting it on a logarithmic scale (see Fig.~\ref{fig:asymptotic}). We typically use $\omega_{N_2}/E_F = 10^3$ and $k_{N_1}/k_F = 5$ to converge to the expected asymptotic behaviour. 

\begin{figure}
    \centering
    \includegraphics[width = \textwidth]{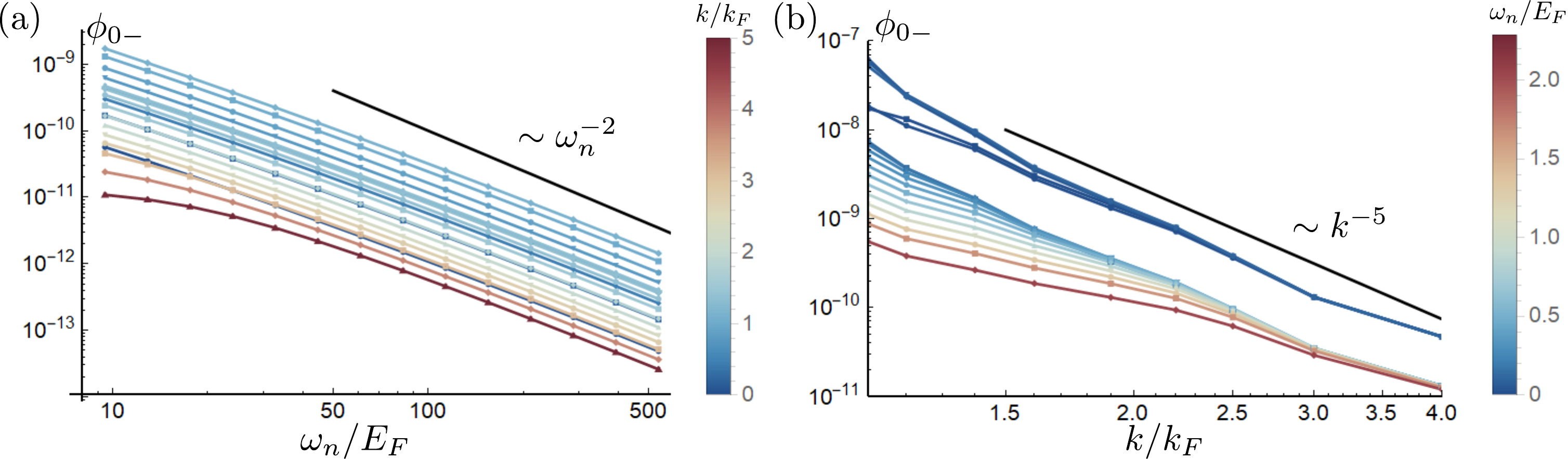}
    \caption{Asymptotic behaviour of the gap function for (a) large frequencies for different $k/k_F$ and for (b) large wavevectors for different $\omega_n/E_F$, for $\ell = 0$ and $r_s = 5$. The black line is a guide to the eye for the expected asymptotic behaviour.}
    \label{fig:asymptotic}
\end{figure}

\section{Random phase approximation}
\label{app:rpa}

The effect of retardation is discussed as for the electron-phonon coupling [\cite{rickayzen}]. We have the equation of motion, for arbitrary $\hat{O}$,
\begin{align}
    &\bigg[ \frac{\partial}{\partial\tau} - \left( \xi_{\sigma_1}({\bf k}) - \xi_{\sigma_2}({\bf k}-{\bf q}) \right)\bigg] \left\langle \hat{T}_{\tau} \hat{c}^{\dagger}_{{\bf k}\sigma_1\lambda_1}(\tau) \hat{c}_{{\bf k}-{\bf q}\sigma_2\lambda_2}(\tau) \hat{O} \right\rangle =\\
    &+\frac{1}{\mathcal{V}}\sum_{{\bf p} {\bf q}^{\prime}\neq0}\sum_{\substack{\sigma_3\sigma_4\sigma_5\\ \lambda_3\lambda_4\lambda_5}}  V_0({ q})  \bigg\langle \hat{T}_{\tau} \bigg( \langle \sigma_3\lambda_3 {\bf k} - {\bf q}|\sigma_1\lambda_1 {\bf k} \rangle \hat{c}^{\dagger}_{{\bf k}- {\bf q}^{\prime}\sigma_3\lambda_3}\hat{c}_{{\bf k}-{\bf q} \sigma_2\lambda_2} - \langle \sigma_2\lambda_2 {\bf k} - {\bf q}|\sigma_3\lambda_3 {\bf k} - {\bf q} + {\bf q}^{\prime} \rangle \hat{c}^{\dagger}_{{\bf k}\sigma_1\lambda_1}\hat{c}_{{\bf k}-{\bf q} + {\bf q}^{\prime}\sigma_3\lambda_3}  \bigg)\nonumber\\
    &~~~~~~~~~~~~~~~~~~~~~~~~~~~~\times\langle \sigma_4\lambda_4 {\bf p} + {\bf q}^{\prime}|\sigma_5\lambda_5 {\bf p} \rangle \hat{c}^{\dagger}_{{\bf p}+{\bf q}^{\prime}\sigma_4\lambda_4}\hat{c}_{{\bf p}\sigma_5\lambda_5} \hat{O} \bigg\rangle\nonumber,
\end{align}
which we simplify in the random phase approximation (RPA) by discarding contributions with ${\bf q} \neq {\bf q}^{\prime}$ and applying a Wick decomposition on the right-hand side. Then
\begin{align}
    &\left[ \frac{\partial}{\partial\tau} - \left( \xi_{\sigma_1}({\bf k}) - \xi_{\sigma_2}({\bf k}-{\bf q}) \right)\right] \left\langle \hat{T}_{\tau} \hat{c}^{\dagger}_{{\bf k}\sigma_1\lambda_1}(\tau) \hat{c}_{{\bf k}-{\bf q}\sigma_2\lambda_2}(\tau) \hat{O} \right\rangle = \\
    &
    \frac{1}{\mathcal{V}}V_0({ q}) \langle \sigma_2 \lambda_2 {\bf k} - {\bf q}|\sigma_1 \lambda_1 {\bf k}\rangle  \left( f_D(\xi_{\sigma_2}({\bf k}-{\bf q})) - f_D(\xi_{\sigma_1}({\bf k})) \right) \sum_{\substack{{\bf p}\sigma_3\sigma_4\\\lambda_3\lambda_4}} \langle \sigma_3\lambda_3 {\bf p}|\sigma_4\lambda_4 {\bf p} - {\bf q} \rangle \left\langle \hat{T}_{\tau} \hat{c}^{\dagger}_{{\bf p}\sigma_3\lambda_3}(\tau) \hat{c}_{{\bf p} - {\bf q}\sigma_4\lambda_4}(\tau) \hat{O} \right\rangle\nonumber,
\end{align}
where $f_D(\xi)$ is the Fermi-Dirac distribution. We perform the decomposition in a contribution independent on the potential and its correction, up to the contribution in $\hat{O}$, $\left\langle \hat{T}_{\tau} \hat{c}^{\dagger}_{{\bf k}\sigma_1\lambda_1}(\tau) \hat{c}_{{\bf k}-{\bf q}\sigma_2\lambda_2}(\tau) \hat{O} \right\rangle = \left\langle \hat{T}_{\tau} \hat{c}^{\dagger}_{{\bf k}\sigma_1\lambda_1}(\tau) \hat{c}_{{\bf k}-{\bf q}\sigma_2\lambda_2}(\tau) \hat{O} \right\rangle_0 + \left\langle \hat{T}_{\tau} \hat{c}^{\dagger}_{{\bf k}\sigma_1\lambda_1}(\tau) \hat{c}_{{\bf k}-{\bf q}\sigma_2\lambda_2}(\tau) \hat{O} \right\rangle_1$. Then, after a Fourier transformation of the equation, multiplying it by a factor $\langle \sigma_1\lambda_1 {\bf k}|\sigma_2\lambda_2 {\bf k} - {\bf q} \rangle$ and summing over $\sigma_1,\lambda_1,\sigma_2,\lambda_2$ and ${\bf k}$ one finds
\begin{align}
    \sum_{\substack{{\bf k}\sigma_1\sigma_2\\\lambda_1\lambda_1}} \langle \sigma_1\lambda_1 {\bf k}|\sigma_2\lambda_2 {\bf k} - {\bf q} \rangle &\left\langle \hat{T}_{\tau} \hat{c}^{\dagger}_{{\bf k}\sigma_1\lambda_1}(i\omega_n) \hat{c}_{{\bf k}-{\bf q}\sigma_2\lambda_2}(i\omega_n) \hat{O} \right\rangle_1\\
    &= V_0({ q}) \Pi(i\omega_n,{\bf q}) \sum_{\substack{{\bf k}\sigma_1\sigma_2\\\lambda_1\lambda_1}} \langle \sigma_1\lambda_1 {\bf k}|\sigma_2\lambda_2 {\bf k} - {\bf q} \rangle \left\langle \hat{T}_{\tau} \hat{c}^{\dagger}_{{\bf k}\sigma_1\lambda_1}(i\omega_n) \hat{c}_{{\bf k}-{\bf q}\sigma_2\lambda_2}(i\omega_n) \hat{O} \right\rangle,\nonumber
\end{align}
with $\Pi(i\omega_n,{\bf q}) = \frac{1}{\mathcal{V}}\sum_{\substack{{\bf p} \sigma_1 \sigma_2\\\lambda_1\lambda_2}} |\langle \sigma_1 \lambda_1 {\bf p}|\sigma_2 \lambda_2 {\bf p }-{\bf q}\rangle|^2 \left( \frac{f_D(\xi_{\sigma_2}({\bf p}-{\bf q})) - f_D(\xi_{\sigma_1}({\bf p}))}{-i\omega_n + \xi_{\sigma_2}({\bf p}-{\bf q}) - \xi_{\sigma_1}({\bf p}) } \right)  $ is the RPA polarisability. The resulting expression is then
\begin{align}
    \sum_{\substack{{\bf k}\sigma_1\sigma_2\\\lambda_1\lambda_1}} \langle \sigma_1\lambda_1 {\bf k}&|\sigma_2\lambda_2 {\bf k} - {\bf q} \rangle \left\langle \hat{T}_{\tau} \hat{c}^{\dagger}_{{\bf k}\sigma_1\lambda_1}(i\omega_n) \hat{c}_{{\bf k}-{\bf q}\sigma_2\lambda_2}(i\omega_n) \hat{O} \right\rangle \\
    &= \frac{1}{1 - V_0({ q}) \Pi(i\omega_n,q)} \sum_{\substack{{\bf k}\sigma_1\sigma_2\\\lambda_1\lambda_1}} \langle \sigma_1\lambda_1 {\bf k}|\sigma_2\lambda_2 {\bf k} - {\bf q} \rangle \left\langle \hat{T}_{\tau} \hat{c}^{\dagger}_{{\bf k}\sigma_1\lambda_1}(i\omega_n) \hat{c}_{{\bf k}-{\bf q}\sigma_2\lambda_2}(i\omega_n) \hat{O} \right\rangle_0\nonumber
\end{align}
and the inverse Fourier transform gives
\begin{align}
    V_{0}({ q})\sum_{\substack{{\bf k}\sigma_1\sigma_2\\\lambda_1\lambda_1}} \langle \sigma_1\lambda_1 {\bf k}&|\sigma_2\lambda_2 {\bf k} - {\bf q} \rangle\left\langle \hat{T}_{\tau} \hat{c}^{\dagger}_{{\bf k}\sigma_1\lambda_1}(\tau) \hat{c}_{{\bf k}-{\bf q}\sigma_2\lambda_2}(\tau) \hat{O} \right\rangle \\
    &= \int d\tau^{\prime} V(\tau - \tau^{\prime},{\bf q}) \sum_{\substack{{\bf k}\sigma_1\sigma_2\\\lambda_1\lambda_1}} \langle \sigma_1\lambda_1 {\bf k}|\sigma_2\lambda_2 {\bf k} - {\bf q} \rangle\left\langle \hat{T}_{\tau} \hat{c}^{\dagger}_{{\bf k}\sigma_1\lambda_1}(\tau^{\prime}) \hat{c}_{{\bf k}-{\bf q}\sigma_2\lambda_2}(\tau^{\prime}) \hat{O} \right\rangle_0,\nonumber
\end{align}
where $V(\tau,{\bf q}) = \sum_{\omega_n} e^{i\omega_n \tau} V_0({ q})/(1-V_{0}({ q})\Pi(i\omega_n,{\bf q}))$ is the retarded Coulomb potential.

\section{Green function integrated over angles}
\label{app:gtheta}

The calculation of the self-energy in Eq.~\eqref{eq:app:self} involves the following integral
\begin{align}
    \Sigma^{(N)}_{\sigma\lambda}(i\omega_n,{\bf p})  &= -(2\beta)^{-1} \sum_{\Omega_m\sigma^{\prime}} \frac{1}{(2\pi)^3} \int dq q^{2} \frac{V_0({q})}{\epsilon(i\Omega_m,{q})}\int d\theta d\phi \sin(\theta) {\rm Tr}\left( \hat{P}_{\sigma}({\bf p}) \hat{P}_{\sigma^{\prime}}({\bf p}-{\bf q}) \right) G^{(N0)}_{\sigma^{\prime}}(i(\omega_n - \Omega_{m}),{\bf p}-{\bf q})\\
     &= \beta^{-1} \sum_{\Omega_m} \int dq q^2 \frac{V_0({q})N_0}{\epsilon(i\Omega_m,{q})}\bar{G}_{\sigma}(i(\omega_n-\Omega_m),p,q)
\end{align}
where we have introduced the averaged coupling function over angles
\begin{align}
    \bar{G}_{\sigma}(i(\omega_n-\Omega_m),p,q) &= \sum_{\sigma^{\prime}}\int_{-1}^{1} du \frac14\left\{ 2 + \sigma \sigma^{\prime} \left[  \frac{3(p + q u)^2}{p^2 + q^2 + 2pq u} - 1\right] \right\}\frac{1}{i(\omega_n-\Omega_m) - (\sigma^{\prime}(p^2+q^2+2p q u) - {\rm sgn}(E_F))}.
\end{align}
We decompose this function over the real and imaginary parts, $\bar{G}^{(1)}$ and $\bar{G}^{(2)}$,
\begin{align}
    \bar{G}_{\sigma}(i(\omega_n-\Omega_m),p,q) = \bar{G}^{(1)}_{\sigma}(i(\omega_n-\Omega_m),p,q) + i\bar{G}^{(2)}_{\sigma}(i(\omega_n-\Omega_m),p,q)
\end{align}
where
\begin{align}
    &\bar{G}^{(1)}_{\sigma}(i(\omega_n-\Omega_m),p,q) = \frac12 \sum_{\sigma^{\prime}} \left\{ -\frac{3\sigma}{4p^2} + \frac{1}{16pq} \frac{1}{\sigma^{\prime} p^2} \frac{3\sigma {\rm sgn}(E_F)(p^2-q^2)^2}{1 + (\Omega_m-\omega_n)^2} \log\left( \frac{(p+q)^2}{(p-q)^2} \right)\right.\\
    &+ \frac{1}{16pq} \frac{1}{\sigma^{\prime} p^2} 3\sigma(\Omega_m-\omega_n)\left(1 - \frac{(p^2-q^2)^2}{1 + (\Omega_m - \omega_n)^2} \right) \left( \arctan\left[ \frac{{\rm sgn}(E_F) - \sigma^{\prime}(p+q)^2}{\omega_n- \Omega_m}\right] - \arctan\left[ \frac{{\rm sgn}(E_F) - \sigma^{\prime}(p-q)^2}{\omega_n- \Omega_m}\right]\right)\nonumber\\
    &+\frac{1}{32 pq}\frac{1}{\sigma^{\prime}p^2}\left( 8p^2 + (3{\rm sgn}(E_F) + 2(p^2-3q^2)\sigma^{\prime})\sigma + \frac{3(p^2-q^2)^2 {\rm sgn}(E_F) \sigma}{1 + (\Omega_m-\omega_n)^2} \right) \log\left( \frac{({\rm sgn}(E_F) - \sigma^{\prime}(p-q)^2)^2 + (\Omega_m-\omega_n)^2}{({\rm sgn}(E_F) - \sigma^{\prime}(p+q)^2)^2 + (\Omega_m-\omega_n)^2} \right) \bigg\}\nonumber,\\
    &\bar{G}^{(2)}_{\sigma}(i(\omega_n-\Omega_m),p,q) = \frac12 \sum_{\sigma^{\prime}} \left\{ \frac{1}{16pq} \frac{1}{\sigma^{\prime} p^2} \frac{3\sigma (p^2-q^2)^2 (\Omega_m-\omega_n)}{1 + (\Omega_m-\omega_n)^2} \log\left( \frac{(p+q)^2}{(p-q)^2} \right) + \frac{1}{16pq} \frac{1}{\sigma^{\prime} p^2} \bigg( 8p^2 + (3{\rm sgn}(E_F) \right.\\
    &\left. + 2(p^2-3q^2)\sigma^{\prime})\sigma + \frac{3(p^2-q^2)^2 {\rm sgn}(E_F) \sigma}{1 + (\Omega_m-\omega_n)^2} \right) \left( \arctan\left[ \frac{{\rm sgn}(E_F) - \sigma^{\prime}(p+q)^2}{\omega_n- \Omega_m}\right] - \arctan\left[ \frac{{\rm sgn}(E_F) - \sigma^{\prime}(p-q)^2}{\omega_n- \Omega_m}\right]\right)\nonumber\\
    &-\frac{1}{32 pq}\frac{1}{\sigma^{\prime}p^2}3\sigma(\Omega_m-\omega_n)\left(1 - \frac{(p^2-q^2)^2}{1 + (\Omega_m - \omega_n)^2} \right) \log\left( \frac{({\rm sgn}(E_F) - \sigma^{\prime}(p-q)^2)^2 + (\Omega_m-\omega_n)^2}{({\rm sgn}(E_F) - \sigma^{\prime}(p+q)^2)^2 + (\Omega_m-\omega_n)^2} \right) \bigg\}\nonumber,
\end{align}
where we have kept track of the sign of the Fermi energy, ${\rm sgn}(E_F)$, in the calculation. All results reported in the main text are obtained for ${\rm sgn}(E_F) = -1$.
\end{widetext}
\end{document}